\documentclass{jfm}
\usepackage{graphicx}
\usepackage{epstopdf, epsfig,color}
\usepackage{amssymb}
\usepackage{wrapfig,txfonts}
\usepackage{natbib}
\usepackage{url}

  \newcommand{\thalf}{\textstyle{\frac{1}{2}} \displaystyle }
   \newcommand{\tthird}{\textstyle{\frac{1}{3}} \displaystyle }
    \newcommand{\twothirds}{\textstyle{\frac{2}{3}} \displaystyle }

   \newcommand{\be}{\begin{equation}}
      \newcommand{\ee}{\end{equation}}
         \newcommand{\tn}{\textnormal}
  \newcommand{\bea}{\begin{eqnarray}}
   \newcommand{\eea}{\end{eqnarray}}

\title[Spreading or contraction of  viscous drops between plates]{Spreading or contraction of  viscous drops between plates: single, multiple or annular drops}

\author[ H.K. Moffatt, Howard Guest  \& Herbert E. Huppert ]{ H. K. Moffatt$^1$, Howard Guest$^2$, Herbert  E. Huppert$^3$ }

\affiliation{$^1$Department of Applied Mathematics and Theoretical Physics, \\
Wilberforce Road, Cambridge CB3 0WA, UK\\
[\affilskip]$^2$ St John's Innovation Centre, Cowley Rd, Milton, Cambridge CB4 0WS\\
[\affilskip]$^3$ Institute of Theoretical Geophysics, King’s College, Cambridge   CB2 1ST}

\date{}
\begin{document}

\maketitle

\begin{abstract}
The behaviour of a viscous drop squeezed between two horizontal planes (a contracting Hele-Shaw cell) is treated by both theory and and experiment.  When the squeezing force $F$ is constant and surface tension is neglected, the theory predicts ultimate growth of the radius $a\sim t^{1/8}$ with time $t$.  This theory is first reviewed and found to be in excellent agreement with experiment. Surface tension at the drop boundary reduces the interior pressure, and this effect is included in the analysis, although it is negligibly small in the squeezing experiments.   An initially elliptic drop tends to become circular as $t$ increases. More generally, the circular evolution is found to be stable under small perturbations.  If, on the other hand, the force is reversed ($F<0$), so that the plates are drawn apart (the `contraction problem'), the boundary of the drop is subject to a fingering instability on a scale determined by surface tension.  The effect of a trapped air bubble at the centre of the drop is then considered.  The annular evolution of the drop under constant squeezing is still found to follow a `one-eighth' power law, but this is unstable, the instability originating at the boundary of the air bubble, i.e.~the inner boundary of the annulus. The air bubble is realised experimentally in two ways:  first  by simply starting with the drop in the form of an annulus, as nearly circular as possible; and second by forcing four initially separate drops to expand and merge, a process that involves the resolution of `contact singularities' by surface tension.  If the plates are drawn apart, the evolution is still subject to the fingering instability driven from the outer boundary of the annulus. This instability is realised experimentally by levering the plates apart at one corner: fingering develops at the outer boundary and spreads rapidly to the interior as the levering is slowly increased.  At a later stage, before ultimate rupture of the film and complete separation of the plates, fingering spreads also from the boundary of any interior trapped air bubble, and small cavitation bubbles appear in the very low pressure region far from the point of leverage. This exotic behaviour is discussed in the light of the foregoing theoretical analysis.
 \end{abstract}
 
 \section{Introduction}
 The behaviour of a drop or film of viscous liquid trapped between two horizontal plates (a Hele-Shaw cell), when the plates are squeezed together by a  force $F(t)$, presents a problem of great practical importance in the adhesion industry.  In this context,  investigations date back to \cite{S1874}, who derived the relation between $F(t)$ and the separation of the plates $h(t)$ when the upper plate is a circular disc that is impelled downwards towards a fixed substrate, and the film extends to the  disc boundary; the same situation was  treated in the more accessible review by \cite{B47}.

 When the drop does not extend to the plate boundaries, and when the squeezing force $F$ is positive and constant (as is the case  in figure \ref{Fig_expanding_drop}), it is known on the basis of lubrication theory that, when surface tension is neglected, the separation  $h(t)\sim t^{-1/4}$ (\citealt{M77}),  and the corresponding drop radius $a(t)$ increases like $a(t)\sim t^{1/8}$ (while the gap contracts, the drop radius expands).  This `one-eighth' power law also applies to the gravitational spread of a viscous liquid on a horizontal plane due to the gradient of the free surface (\citealt{H1982}), with important implications for the spread of  lava from volcanic eruptions (Huppert et al.~1982).  For the squeezing problem, the effect of surface tension, which provides a pressure jump across the air-liquid interface, has been considered by \cite{W06} (see \S\ref{The squeezing problem} below, and also  \S\ref{Sec_stability} where stability of the basic state is considered). Surface tension is of critical importance when the drop is considered as forming a `liquid bridge' between the two plates (see, for example, \citealt{G96}).
  \begin{figure}
\begin{center}
\includegraphics*[width=0.8\textwidth,  trim=0mm 210mm 0mm 0mm]{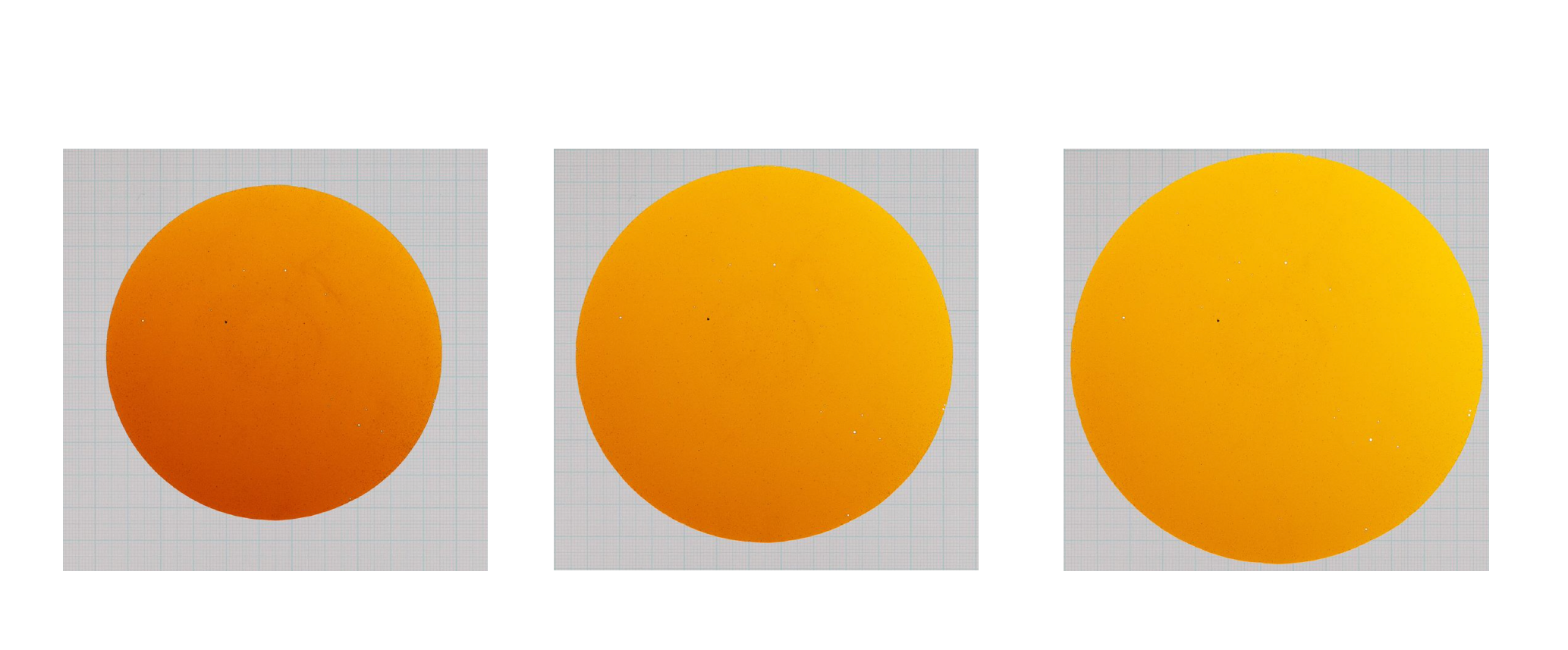} \\
(a) $\tau=60$ \qquad\qquad \qquad(b) $\tau=172$ \qquad\qquad \qquad (c)  $\tau=321$
\end{center}
\caption{Photographs of an expanding viscous drop placed between two horizontal glass plates, the upper plate being allowed to descend under its own weight; the drop is illuminated from below and viewed from above, and the shots were taken at  the dimensionless times $\tau=t/t_{0}$ indicated [here $t_{0}= 2.15$s, defined by (\ref{t_zero})].  The background grid (which may be seen by zooming to expand the figure) covers a 130 mm square in each panel, and allows measurement of the radius $a(\tau)$ to within $\pm 0.1$mm. Note how the opacity decreases as the layer thickness $h(\tau)$ decreases.}  
  \label{Fig_expanding_drop}
\end{figure} 
   \begin{figure}
\begin{center}
\includegraphics*[width=1.1\textwidth,  trim=10mm 0mm 0mm 0mm]{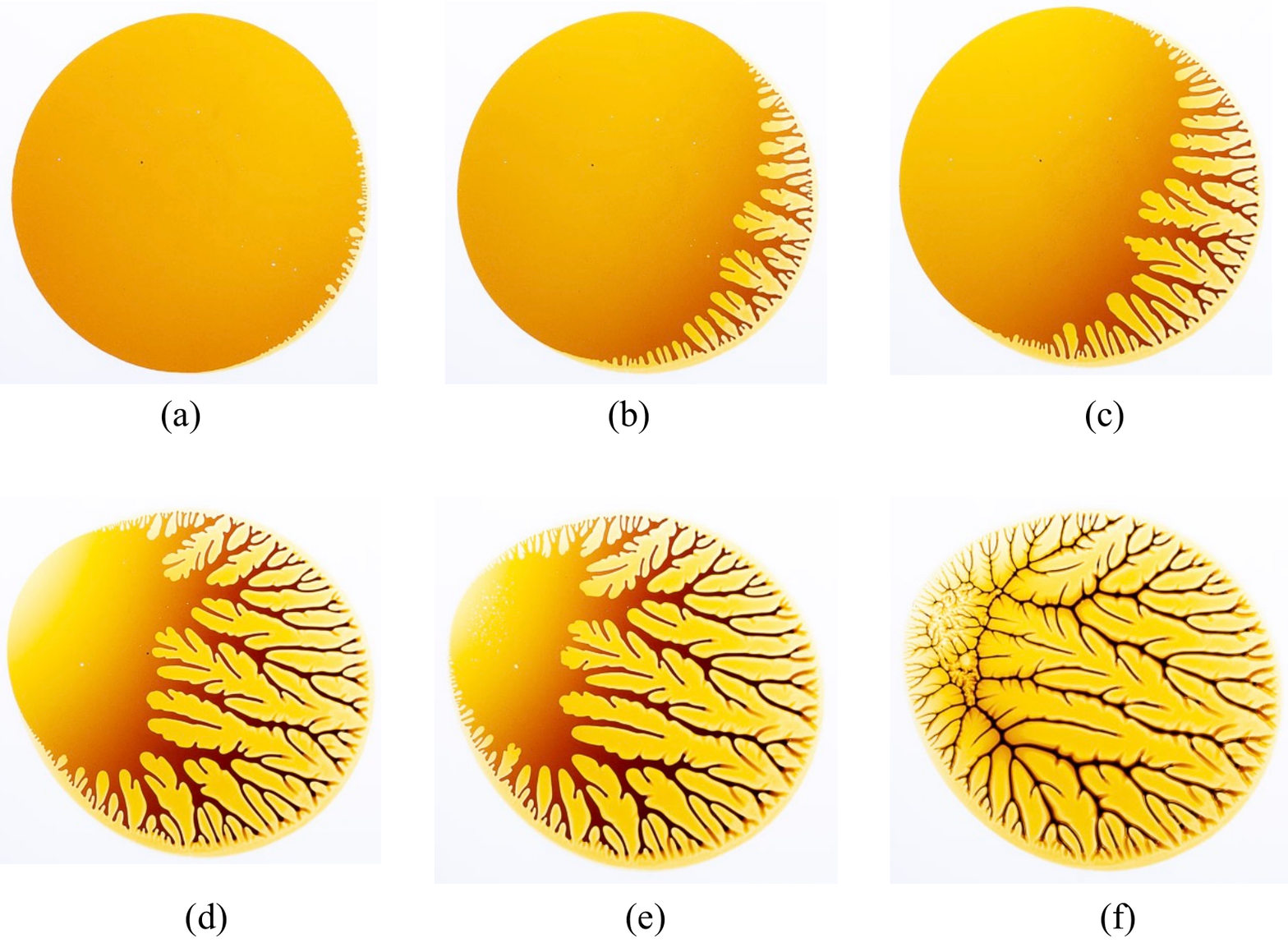} \\
\end{center}
\vskip -110mm
\caption{Continuation of figure \ref{Fig_expanding_drop}, with levering at lower right-hand corner, in order to gradually open the gap between the plates; the leverage here was increased slowly for about one minute, before ultimate rupture of the film; (a) early stage of fingering instability which starts from the `south-east' sector of the boundary; (b) tip-splitting and side-branching begins as fingers grow; (c-e) the process continues, and in (e) cavitation bubbles can be detected in the `north-west' area of the drop; (f) fingering fills the fluid domain, and a `ridge' is evident where fingers emanating from the north-west sector of the boundary impact those from the south-east.}  
  \label{Fig_Single_drop_fingering}
\end{figure} 

 When the force $F$ is negative, i.e.~when the plates are pulled apart with initial conditions $a(0)=a_{0},\,h(0)=h_{0}\ll a_{0}$,   then (see \S\ref{Sec_contraction_problem}) lubrication theory leads to the results $h(t)=h_{0}(1-t/|t_{0}|)^{-1/4}$ and $a(t)=a_{0}(1-t/|t_{0}|)^{1/8}$, where $|t_{0}|$ is a characteristic time proportional to $|F|/\mu$ and $\mu$ is the dynamic viscosity of the fluid. This indicates an incipient `finite-time singularity' at time $t=|t_{0}|$ when, in effect, the plates separate completely; but lubrication theory ceases to be valid when the critical time is approached, specifically when 
 $1-t /|t_{0}|\sim(h_{0}/a_{0})^{8/3}$.

There are additional complications when $F\!<\!0$, which have been reviewed by  \cite{G02}. First, there is an instability due to the suction of air into the region occupied by the more viscous liquid as its area decreases (see figure \ref{Fig_Single_drop_fingering}, which shows the effect of levering the plates apart at one corner).  This is a fingering  instability, as first described by \cite{S58}, and  as encountered in the Hele-Shaw experiments of \cite{R88}, different only in that in these experiments air was injected at high pressure into the lower-pressure viscous layer.  Following the computational investigation of  \cite {K97},  it was shown by \cite{T2000} that the stability problem is ill-posed in this sort of situation, but that the potential development of singularities can be controlled by surface tension; this is consistent with the analysis of the fingering instability that we present in \S\ref{Sec_fingering_instability}. Fractal patterns generated by fingering in a separating-plate experiment have been compared with the results of  statistical simulations by  \cite{L91}. We note that fingering can be  controlled to some extent, not only by surface tension, but also through replacement of the plates 
by elastic membranes, a situation of potential importance in biological contexts (\citealt{P12}).   

Secondly, cavitation bubbles may appear in the course of plate separation, presumably wherever the pressure would otherwise fall below the vapour pressure of the viscous liquid;  cavitation bubbles may be detected in figure \ref{Fig_Single_drop_fingering}(e). a phenomenon studied in detail by  \cite{L99} using a `probe-tack apparatus'  and acrylic pressure-sensitive adhesives.   When varying the traction velocity, there is a transition from a fingering regime to a cavitation regime, which has been investigated by \cite{P04}.  Again, the effect of surface tension, which can counteract the traction force when the gap is sufficiently small and is therefore crucial for effective adhesion, has been treated by  \cite{W11}. This also will be reviewed in \S\ref{Sec_contraction_problem}, and  again in the stability analysis of \S\ref{Sec_fingering_instability}.
\begin{figure}
\begin{center}
\includegraphics*[width=1\textwidth,  trim=0mm 0mm 0mm 5mm]{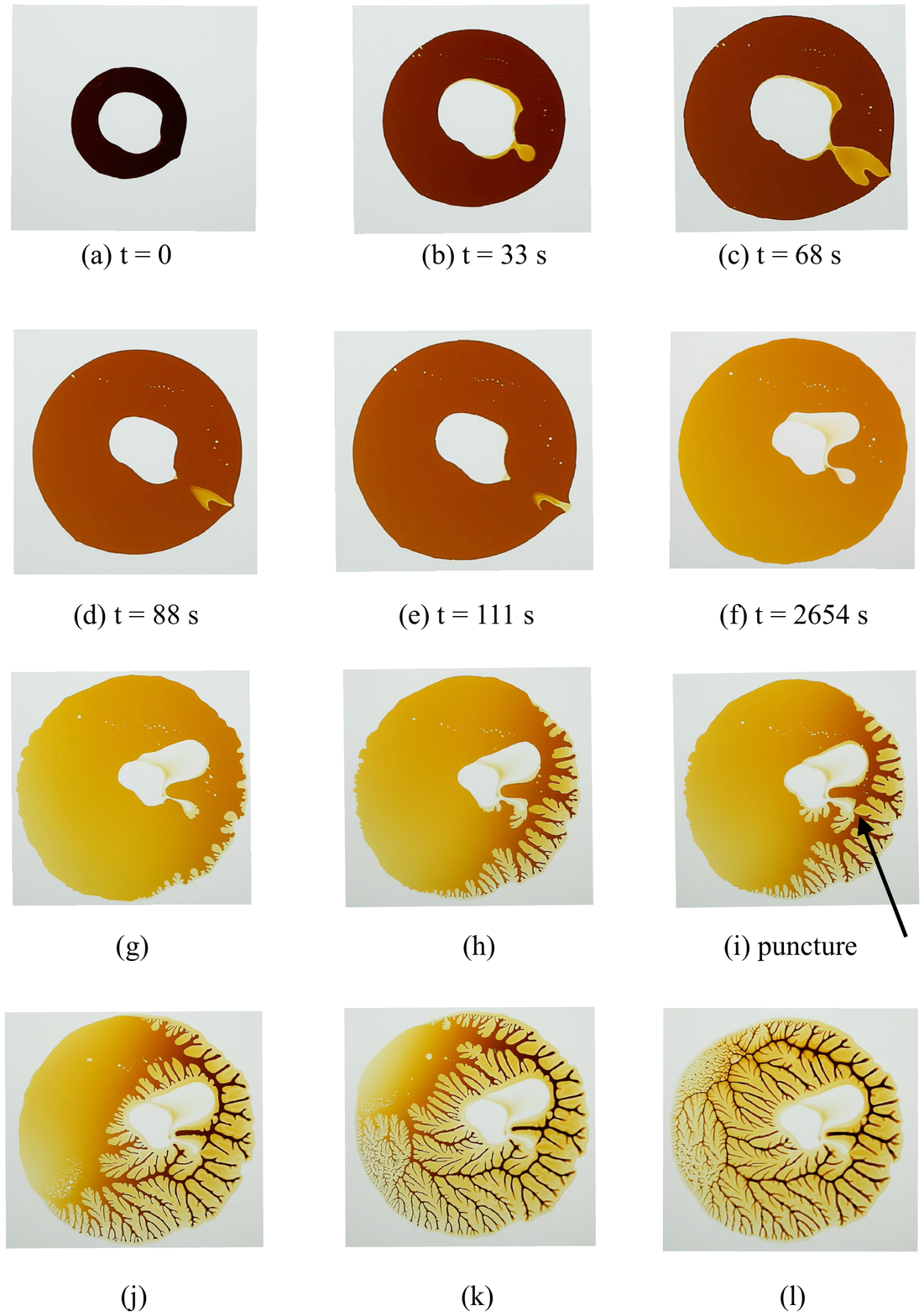} 
\end{center}
\vskip -20mm
\caption{Evolution of an annular drop;  stills at time $t$ (seconds) as indicated for panels (a-f). Panel (a): the initial, nearly circular, annular drop; (b) eruption of a secondary bubble from where the internal curvature is maximal; the outer boundary is only weakly perturbed; (c) tip-splitting of the secondary bubble, which has just penetrated the outer boundary; note that the colour shows that a very thin `wetting' layer of treacle has been left on both plates where the secondary bubble has evolved; (d) eruption of the secondary bubble to the exterior as it detaches from the primary bubble; (e) eruption nearly complete; (f) eruption of tertiary bubbles from the primary bubble; (g) levering at lower right-hand corner of the plates is introduced, showing early development of fingering instability from the outer boundary; (h) the fingers advance showing significant tip-splitting and side-branching; puncturing is imminent; (i) a finger punctures the tertiary bubble (at the point marked by the arrow) allowing equalisation of pressure  with the exterior; (j) no further puncturing occurs, but a ridge is established, separating  the `invading' and `defending' cohorts of fingers; (k) fingering  spreads right round the outer boundary and cavitation bubbles appear ahead of the advancing front; (l) the fingering pattern is now fully established, and the ridge remains prominent, with a small gap only at the puncture point.}
  \label{Fig_Annulus}
\end{figure}
\begin{figure}
\begin{center}
\includegraphics*[width=.8\textwidth,  trim=20mm 80mm 20mm 10mm]{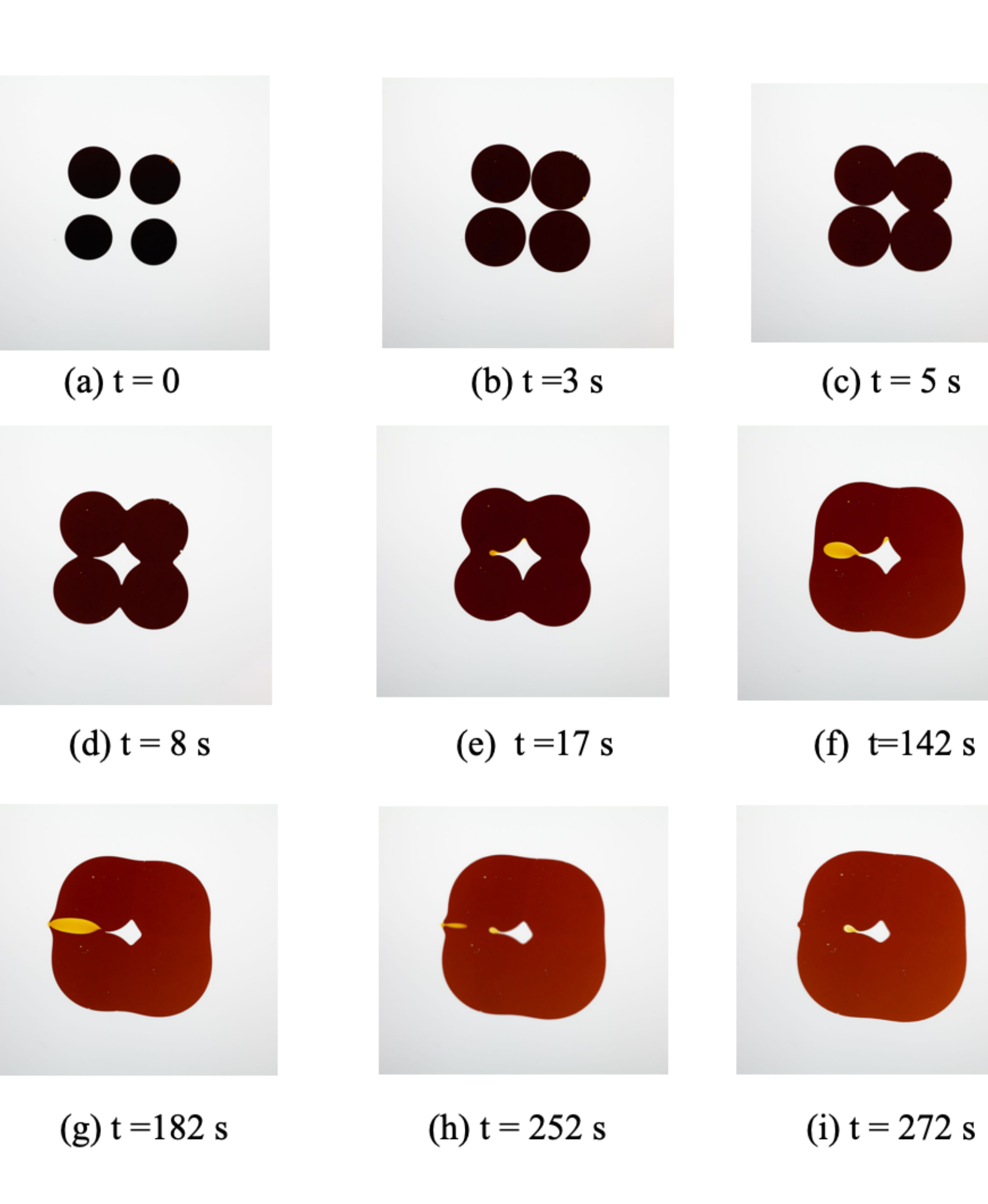} 
\end{center}
\caption{Expansion and merging of four drops of treacle between two horizontal plates; time $t$ in seconds as indicated.  An air bubble is trapped, but does not relax to circular form; the volume of air in the bubble (or bubbles)  is constant for so long as they remain trapped. Panel (a): initial configuration; (b) the four drops have expanded, and a first contact is imminent; (c) first two contacts have occurred almost simultaneously, forming  cusps which immediately resolve themselves through the effect of surface tension; a third contact is imminent; (d) the air bubble is now trapped, and its volume remains constant; (e) the fourth cusp is resolved by air pushing into the surrounding treacle forming a secondary bubble, while the combined volume of primary and secondary bubbles remains constant; (f) the secondary bubble expands and (g) penetrates the outer boundary; (h) the secondary bubble drains to the exterior, and a tertiary bubble begins to form from the residual cusp on the primary bubble; (i) the secondary bubble has escaped and the tertiary bubble grows very slowly. }
  \label{Fig_four_drops}
\end{figure} 

\begin{figure}
\begin{center}
\includegraphics*[width=1\textwidth,  trim= 0mm 85mm 0mm 10mm]{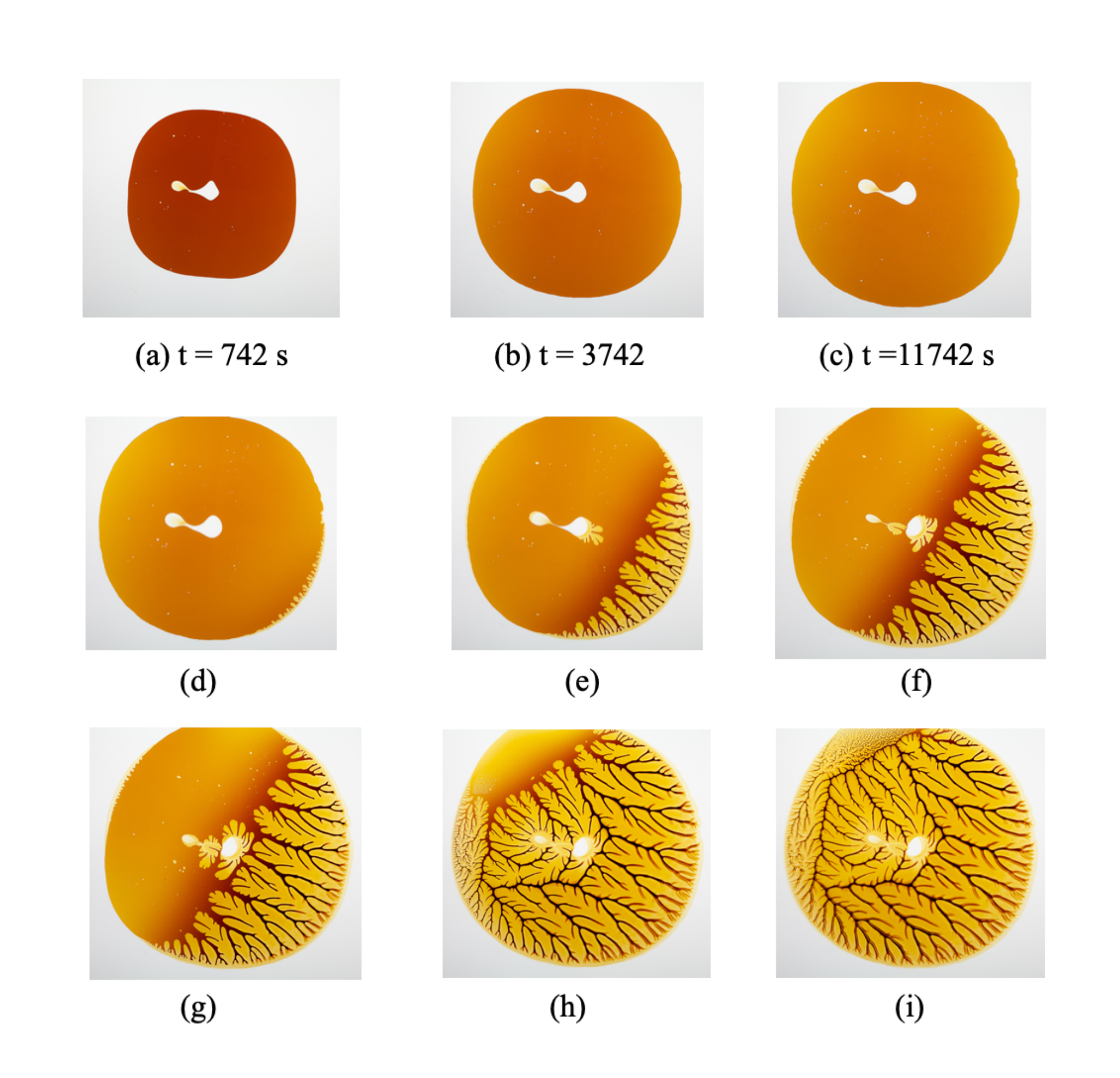} 
\end{center}
\caption{Continuation of figure \ref{Fig_four_drops}.  Panels (a-c): the tertiary bubble grows extremely slowly for over three hours and appears to attain a quasi-equilibrium. Panels (d-i): development of fingering instability induced by  levering the plates apart with a knife inserted at the lower right-hand corner; the levering was gradually increased for about one minute before final rupture;  (d) early stage of fingering instability at the nearly circular outer boundary of the drop; (e) early stage of tip-splitting; (f) early stage of side-branching and growth of `defending' fingers from the  bubble boundary; (g) advanced stage of fingering while the defending fingers retreat and one of the `invading' fingers punctures a defending finger providing a pressure-equalising path from the bubble to the exterior; (g) invading fingers advance round the boundary, and cavitation bubbles appear ahead of the advancing front;  (i) ultimate stage of fingering, when the treacle is confined to the narrow tree-branches and twigs that now separate the fingers. Note here the near-pentagonal ridge that separates the invading fingers from the defending fingers, with again a single breach in the south-east sector of the ridge.}
  \label{Fig_fingering}
\end{figure}

 There have been a number of investigations of the behaviour when the initial shape of the drop is non-circular.   This problem has been previously treated by \cite{S2015} (see also \citealt{S2021}) by means of conformal mapping, and using the Schwarz function (\citealt{D74}, \citealt{V06}) as earlier proposed for the two-phase (Muskat) problem by \cite{H2000}.  In this way, they found a solution for the case of an initially elliptic drop (see \S \ref{Sec_elliptic_drop} for a simpler treatment of this particular problem).   \cite{S2015} also treated the problem of a nearly circular hypotrochoid of three-fold symmetry, and came to the surprising conclusion that, under squeezing and neglecting surface tension, the drop shape `evolves towards a deltoid', a curve with three cusps. This goes against physical intuition, and our stability analysis in \S\ref{linearised_stab} comes to the opposite conclusion, namely, that quite generally, under squeezing with a constant force, any perturbation from the circular shape is eliminated as the area of the drop increases. This behaviour is consistent with the experiments described in \S\ref{Sec_exp}.

When an air bubble is trapped, either accidentally or deliberately, inside the viscous drop, new problems arise.  When the viscous liquid has the form of an annulus trapping a circular air bubble, the `basic-state' evolution can be easily determined and still follows a one-eighth power law evolution (\S\ref{air_bubble});  but we shall find that this basic state, even under squeezing, is unstable, the instability being driven (in a sense that will be quantified) from the bubble boundary (\S\ref{Sec_destabilising_bubble}). For the case of contraction ($F\!<\!0$), we shall find that the fingering instability persists and is driven primarily from the outer boundary.

The details of the experiments providing the photographs of Figures \ref{Fig_expanding_drop}-\ref{Fig_fingering} will be described in \S\ref{Sec_exp}. These home experiments, conducted under lockdown conditions, replicate lecture-room demonstrations that we have used over many years to illustrate the one-eighth expansion law when $F\!>\!0$ and the fingering instability when $F\!<\!0$.  We used Lyle's black treacle as the working fluid, which allows for good colour contrast.  When a single drop of this fluid is placed on a fixed horizontal glass plate, and a second glass plate is placed on the drop and allowed to descend under its own weight, the drop expands and soon adopts an accurately circular form; the slow expansion can be allowed to continue for several hours.  The situation $F\!<\!0$ can then be achieved by gently levering the plates apart at one corner.  The separation $h({\bf x}, t)$ then varies linearly in ${\bf x}=(x,y)$, but the fingering instability, non-uniform around the drop boundary, is nevertheless well illustrated (figure \ref{Fig_Single_drop_fingering}).  Furthermore, cavitation bubbles can be observed where the pressure is minimal, and explained at least qualitatively in terms of the `hinged plate' problem described in \S2 of \cite{M64}.

The trapping of a large air bubble was achieved experimentally in two quite different ways, first by starting with a drop in the form of a circular annulus  (see figure \ref{Fig_Annulus}), and second, by starting with four disjoint drops (figures  \ref{Fig_four_drops} and  \ref{Fig_fingering}), and forcing these to expand under the weight of the upper plate so that they merge, trapping an air bubble that is very non-circular from the outset. In both cases, the instability that is predicted theoretically manifests itself through the ejection of a secondary bubble from the primary bubble, allowing some air to escape.  The subsequent evolution is extremely slow, but the boundary of the air bubble remains very non-circular throughout.  When the plates are levered apart from one corner, the fingering instability appears first from the external boundary, and later from the internal bubble boundary, and a ridge develops, separating these two fingering `cohorts'.

  \section{The basic state}\label{basic_state}
  \subsection{The squeezing problem $(F\!>\!0)$}\label{The squeezing problem}
 Consider what happens when a drop of viscous liquid of volume $V$ and dynamic viscosity $\mu$ is placed between two  horizontal plates $z=0$ and $z=h(t)>0$, the upper plate being impelled towards the fixed lower plate by a constant force $F$.\, It is observed in this situation that, when viewed from above, the drop tends to become of  circular form of expanding radius $r=a(t)$, where, with $h(0)=h_{0}, a(0)=a_{0}$, by conservation of volume, 
  \be\label{volume}
  \pi\, a^{2}(t)\,h(t) =\pi\, a_{0}^{2}\,h_{0}=V= \tn{const.}
  \ee
 We shall describe this evolving circular situation, shown in figure \ref{Fig_expanding_drop}, as the basic state.  In this state, provided $h(t)\! \ll \!a(t)$, standard lubrication theory implies  that (i) the pressure $p(r,t)$ is independent of $z$ and (ii) the radial velocity component $u(r,z,t)$ satisfies $|\partial u/\partial z| \gg |\partial u/\partial r|$ and is driven by the pressure gradient according to the equation
 \be
\frac{ \partial p}{\partial r }= \mu \frac{\partial^{2}u}{\partial \,z\, ^{2}}\,,
\ee
with boundary conditions $u(r,0,t)=u${\large (}$r, h(t), t${\large )}$=0$. Hence
\be
u(r,z,t)=\frac{1}{2\mu}\frac{ \partial p}{\partial r }\,z\,(z-h)\,, 
\ee
and, averaging over $z$,
\be\label{barur}
\bar{u}(r,t) \equiv \frac{1}{h}\int_{0}^{h}u(r,z,t)\, \tn{d}z=-\frac{h^{2}}{12\mu}\, \frac{ \partial p}{\partial r }\,.
\ee
Mass conservation and incompressibility indicate that the radial flux $Q(r,t)=2\pi r\, h\, \bar{u}(r,t) $ at each radius $r$ is equal to the flux $-\pi r^2\, \tn{d}h/\tn{d}t$ driven downwards by the upper plate. It follows that
\be\label{p_of_r}
 \frac{ \partial p}{\partial r}= \frac{6\,\mu\, r}{h^3}\, \frac{\tn{d}h}{\tn{d}t}
\ee
and we  note that
\be\label{Reynolds}
\nabla^{2}p=\frac{1}{r}\frac{\partial}{\partial r} r\, \frac{ \partial p}{\partial r} = \frac{12\mu}{h^3}\, \frac{\tn{d}h}{\tn{d}t}\,.
\ee
This is the particular form of `Reynolds' equation' applicable to this problem (see for example \S 3.6 of \citealt{M77}).

Let $\gamma$ denote the surface tension at the liquid/air boundary.  Assuming a circular meniscus with zero contact angle, there is then a jump of pressure $\gamma/h(t)$ with corresponding suction across this boundary. The boundary condition is then
\be\label{bc_surface}
p=p_{a}-\gamma/h(t) \quad \tn{at}\quad r=a(t)\,,
\ee
where $p_{a}$ is the atmospheric pressure.
 With this boundary condition the pressure in the drop is then given by
\be\label{pressure0}
p(r,t)-p_{a}= \frac{3\mu}{h^{3}(t)}\frac{\tn{d}h}{\tn{d}t}\left(r^2 -a^{2}(t)\right)  - \frac{\gamma}{h(t)}= p_{0}(r,t),\,\, \tn{say}\,.
\ee 
The total vertical force exerted by the fluid on the upper plate is then
\be\label{Fp}
F\!_{p} =2\pi\int_{0}^{a} p_{0}(r,t)\,r \,\tn{d}r = -\frac{3\pi\,\mu}{2h^3}\frac{\tn{d}h}{\tn{d}t}\, a^{4}-\frac{\pi\gamma}{h(t)}a^2\,.
\ee  
Using (\ref{volume}) and neglecting plate inertia, the force balance $F=F\!_{p} $, we then find 
\be\label{force_balance}
F=\frac{3\mu V^2}{8\pi}\frac{\tn{d}}{\tn{d}t}\left(\frac{1}{h^4}\right)-\frac{\gamma V}{h^2}\,.
\ee
With  dimensionless variables
\be\label{t_zero}
X=h_{0}^{2}/h^{2}(t) =a^{4}(t)/a_{0}^{4}\quad\,\,\tn{and}\,\,\ \tau=t/t_{0}\,, \quad \tn{where} \quad t_0=\frac{3}{8\pi} \frac{\mu}{F}\,\frac{V^2}{h_0^4}  = \frac{3\pi^3}{8}\frac{\mu}{F}\frac{a_{0}^{8}}{V^{2}}\,,
\ee
equation (\ref{force_balance}) takes the convenient  dimensionless form
\be\label{X_eqn}
\tn{d}X^{2}/\tn{d}\tau \equiv 2X\,\tn{d}X/\tn{d}\tau  =1+\lambda\, X\quad\tn{with}\quad X(0)=1, 
\ee
where
\be
 \lambda =(\gamma/F) \left(V/h_{0}^2\right)=(\gamma/F) \left(\pi^{2} a_{0}^{4}/V\right)\,.
\ee
Here $\lambda$ is  the relevant dimensionless measure of surface tension. It is small $(\sim 0.003)$ in the experiments described in \S\ref{Sec_exp}, but surface tension effects can evidently still become important as and when $X$ increases to O$\left(\lambda^{-1}\right)$, i.e.~when $a/a_{0}$ increases to O$\left(\lambda^{-1/4}\right)$. 
\begin{figure}
\vskip -20mm
\begin{center}
\includegraphics*[width=0.45\textwidth,  trim=0mm 0mm 0mm 0mm]{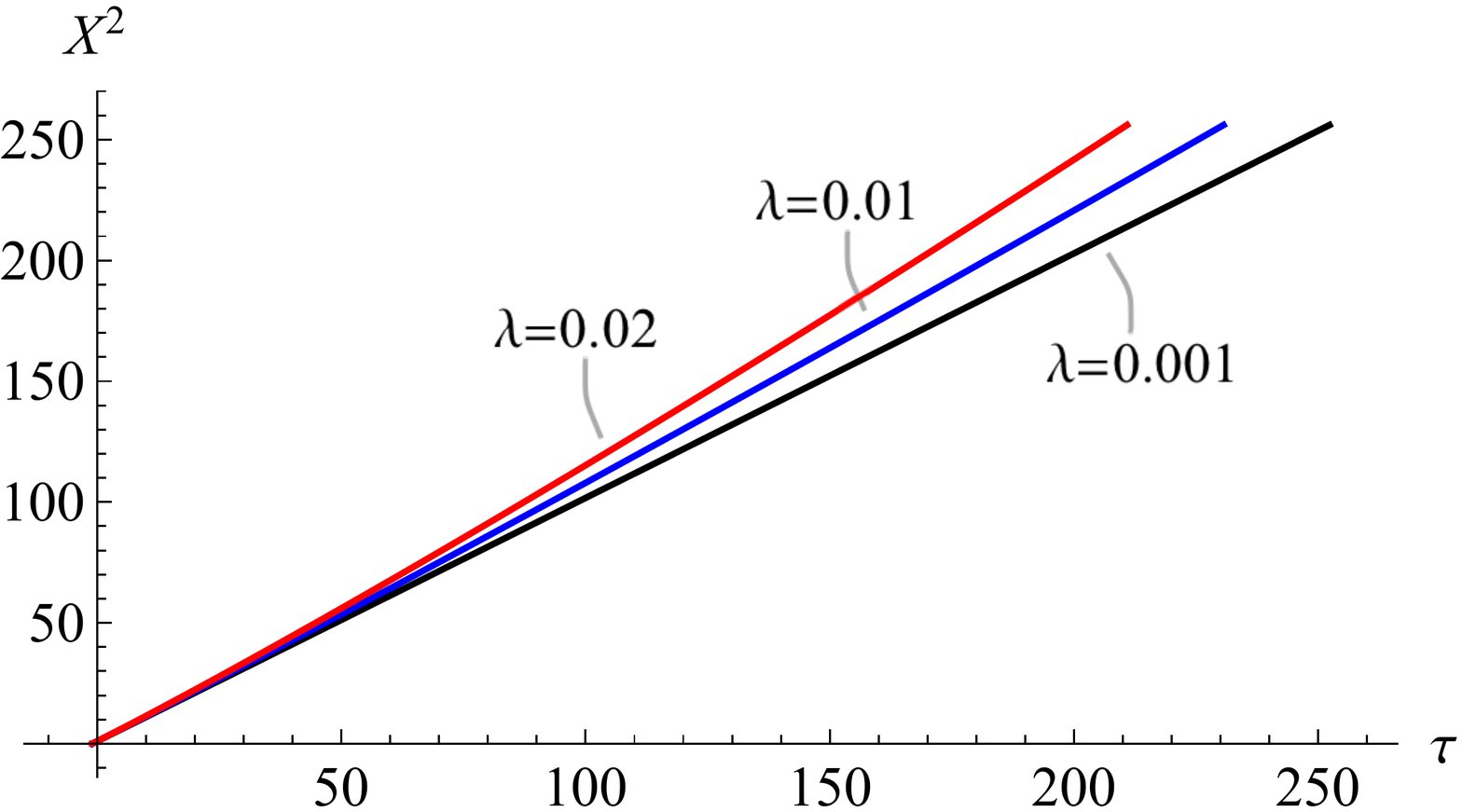} \quad
\includegraphics*[width=0.45\textwidth,  trim=0mm 0mm 0mm 0mm]{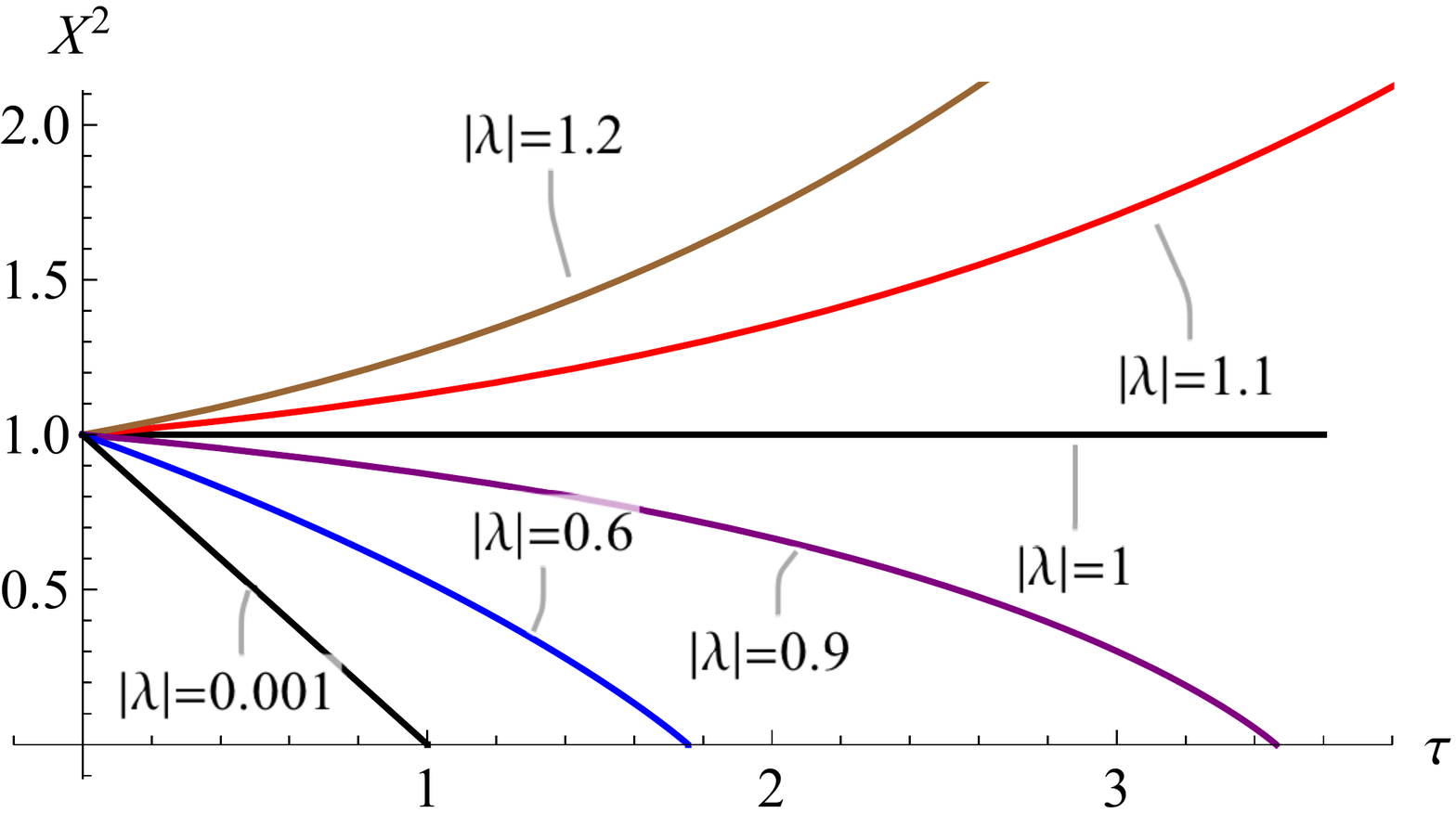} \\
\vskip -22mm
(a)\qquad\qquad\qquad\qquad \qquad\qquad\qquad\qquad\quad (b)
  \end{center}
\caption{(a) Plot of $X^{2}(\tau; \lambda)=a^{8}(t)/a_{0}^{8}$, with  $\tau\equiv t/t_{0}$, as determined by (\ref{t_zero}) and (\ref{implicit_X}), for $\lambda=0.001,0.01,0.02$; note that $X^{2}\sim 1+\tau$ as $\lambda\rightarrow 0$; (b) plot of $X^{2}(\tau; |\lambda|)$, with $\tau=t/|t_{0}|$, for the case of contraction $F<0$ [as given by (\ref{implicit_X_2})]; here $X^{2}\sim 1-\tau$ as $\lambda\rightarrow 0$,  }  
  \label{X_squared}
\end{figure}
\begin{figure}
\begin{center}
\includegraphics*[width=0.45\textwidth,  trim=40mm 0mm 40mm 0mm]{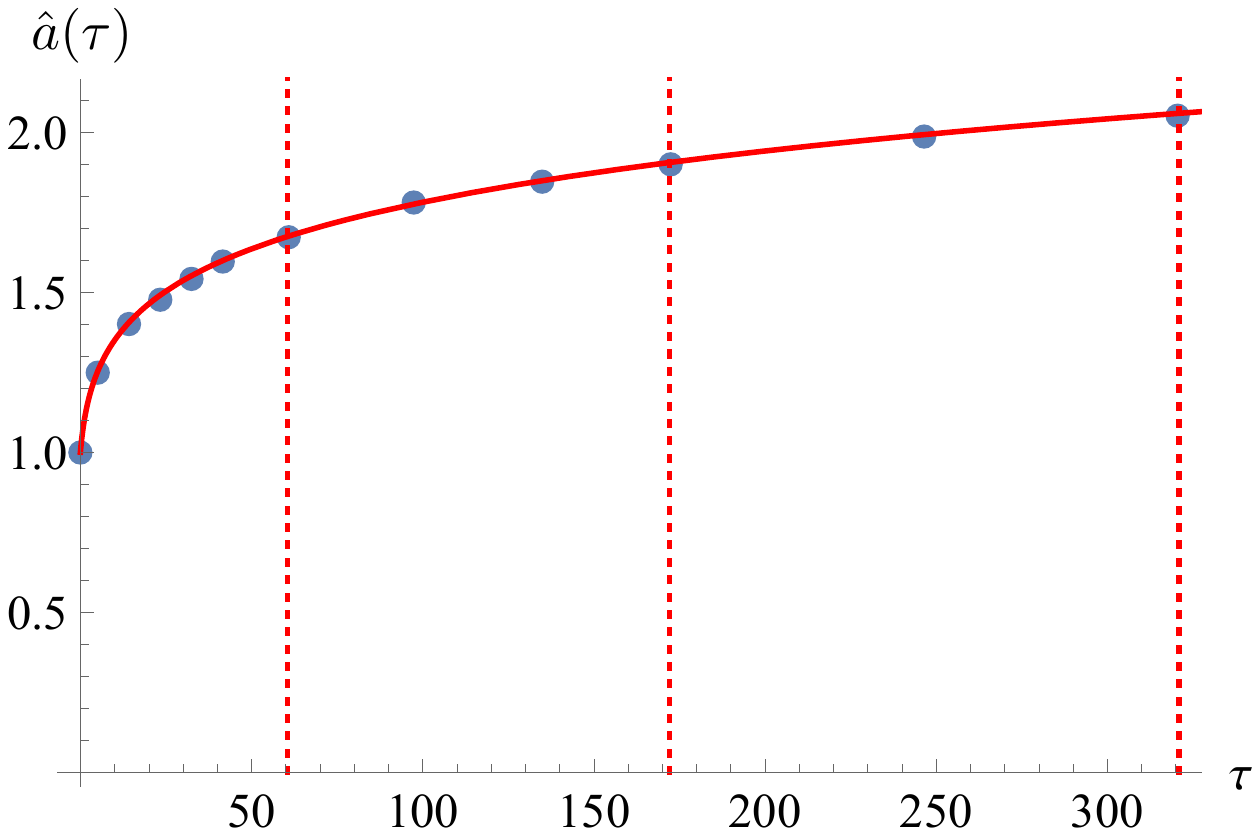} \quad
\includegraphics*[width=0.45\textwidth,  trim=40mm 0mm 40mm 0mm]{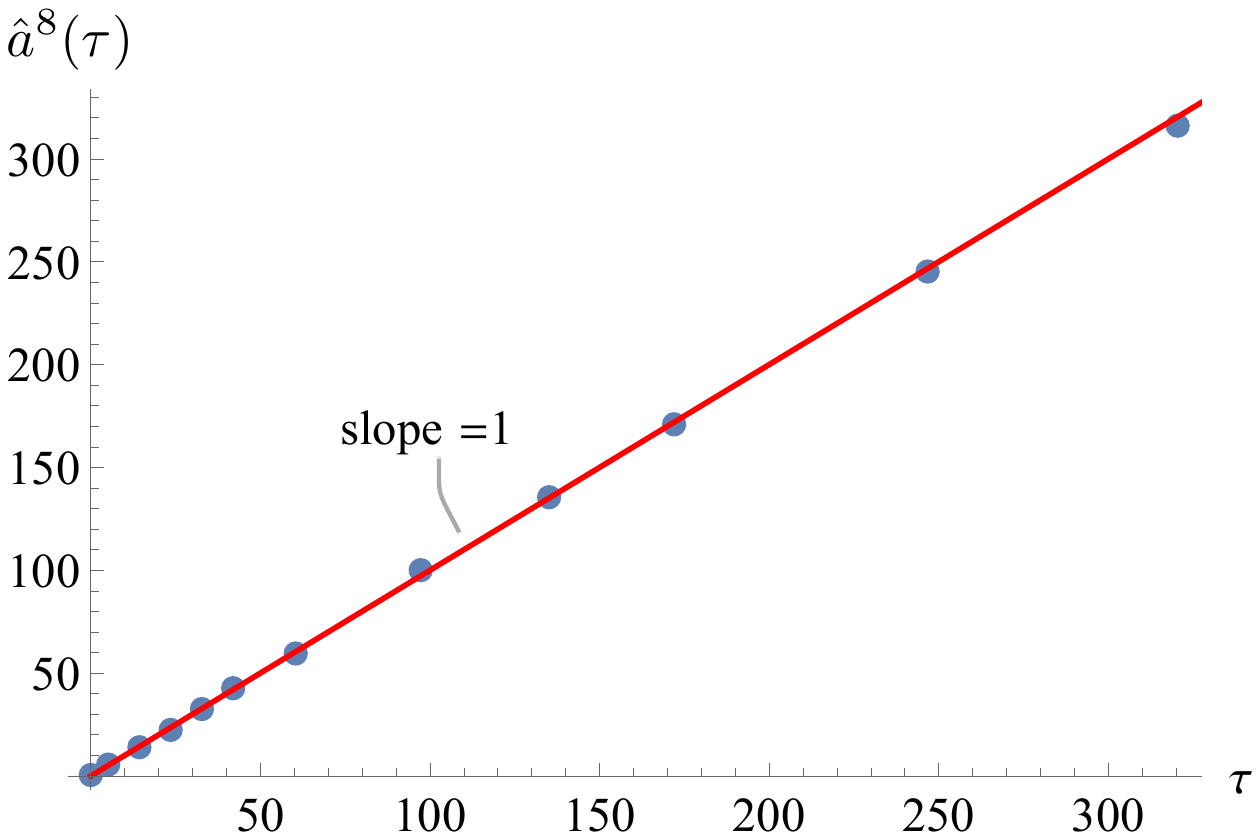} 
\end{center}
\vskip -87mm
\quad\qquad\qquad\qquad\qquad\qquad(a)\qquad\qquad\qquad\qquad\qquad\qquad \qquad\qquad\qquad\qquad(b)
\caption{ (a) Plot of $\hat{a}(\tau)\equiv a(t)/a_{0}$ for a squeezed drop; the red curve shows the function defined by  (\ref{hoft}); the dotted lines correspond to the three images shown in figure \ref{Fig_expanding_drop}; (b) plot of $\hat{a}^8(\tau)$, showing clearly the one-eighth power-law over the full range of $\tau$.}
  \label{Fig_one-eighth}
\end{figure}

Equation (\ref{X_eqn})  integrates to give 
\be\label{implicit_X}
\tau=\frac{2}{\lambda^2} \left((X-1)\lambda -\log{\frac{1+\lambda X}{1+\lambda}}\right)\,,
\ee
thus determining $X$ implicitly as a function of $\tau$ and $\lambda$ (cf.~\citealt{W06}).  For $\lambda\ll 1$, this gives
\be\label{asym_X}
\tau=  X^2-1 - \twothirds \left( X^{3} -1\right) \lambda +\tn{O}\left(\lambda^2\right)\,.
\ee
Figure \ref{X_squared}(a) shows $X^{2}(\tau)$ as determined by (\ref{implicit_X}) for three small values of $\lambda$; as expected from (\ref{asym_X}), the curves all asymptote to $X^{2}\sim 1+\tau$ as $\tau\rightarrow 0$.

If surface tension is neglected (i.e.~$\lambda=0$), then (\ref{asym_X}) gives $X(\tau)=(1+\tau)^{1/2}$ or equivalently
 \be\label{hoft}
 h(\tau)= h_{0}(1+\tau)^{-1/4}\,,\quad  a(\tau)=a_{0}(1+\tau)^{1/8}\,.
 \ee  
The result for $a(\tau)$ is the `one-eighth power law' (as indicated in the introduction).  Figures \ref{Fig_expanding_drop} and \ref{Fig_one-eighth} show the results of a simple experiment (details in \S \ref{Sec_exp});  the one-eighth power-law  shows up quite accurately in this experiment, for which $t_{0}\approx 2.15$s and $\lambda \approx 3\times10^{-3}$.  The time taken to approach this  one-eighth power-law is evidently very short  in this experiment.  This 'adjustment time' has been recently investigated by \cite{B19} and \cite{W20} for the case of gravitational spread of a viscous fluid (\citealt{H1982}). They have found that this  time, somewhat independent of initial conditions, can be just fractions of a second in laboratory experiments, but many days in geological situations, where the extremely viscous lava domes from volcanic eruptions can eventually spread horizontally to hundreds of kilometres.

\subsection{The contraction problem $F<0$}\label{Sec_contraction_problem}
 If $F<0$ (i.e.~if the upper plate is pulled away from the lower plate with a constant traction force $|F|$) leading to contraction of the area of the drop, then $t_{0}<0$, and it is natural therefore to redefine $\tau$
 as $\tau=t/|t_{0}|$; with this modification, (\ref{X_eqn}) becomes
 \be\label{X_eqn_2}
\tn{d}X^{2}/\tn{d}\tau \equiv 2X\,\tn{d}X/\tn{d}\tau  =-1+|\lambda|\, X\quad\tn{with again}\quad X(0)=1.
\ee
Now a steady state with $X=1$ is possible if $|\lambda|=1$; but, if $|\lambda|\ne 1$,
  \be\label{X_eqn_2}
\left|\tn{d}X^{2}/\tn{d}\tau\right|_{\tau=0}   =-1+|\lambda|\,,
\ee
so that  $X^{2}$ increases or decreases from 1 according as $|\lambda| \gtrless 1$.  If $|\lambda| >1$, the reduction in pressure in the drop due to surface tension is sufficient to overcome the traction force, thus decreasing $h(\tau)$ and increasing $a(\tau)$,  ensuring good adhesion.

 Equation (\ref{X_eqn_2}) now integrates to give
\be\label{implicit_X_2}
\tau=\frac{2}{|\lambda|^2} \left((X-1)|\lambda|+\log{\frac{1-|\lambda| X}{1-|\lambda|}}\right)\,,
\ee
again determining $X$ implicitly as a function of $\tau$. Figure \ref{X_squared}(b) shows $X^{2}(\tau)$ as determined by (\ref{implicit_X_2}) for several values of $|\lambda|$, both less than and greater than 1.  For $|\lambda|<1$, $X^2$ goes to zero at a time $\tau^{*}(|\lambda|)$, indicating a finite-time singularity [$h(\tau)=\infty$] at this time.  If $|\lambda|=0$, the solution is
\be\label{implicit_X_2_0}
X=(1-\tau)^{1/2}\quad \tn{so}\quad   h(\tau)=h_{0}\left(1-\tau\right)^{-1/4}\,,\quad  a(\tau)=a_{0} (1- \tau)^{1/8}\,,
\ee 
 indicating the finite-time singularity at $\tau=1$ anticipated in the introduction.
These results hold only for so long as $h(\tau)\ll a(\tau)$, and cease to apply as $\tau=t/|t_{0}|$ approaches the singularity time, specifically when
$1-\tau\sim (h_{0}/a_{0})^{8/3}$; at this stage, it is to be expected that the two plates will  rapidly separate. 
However, as will be shown in \S\ref{Sec_stability} below, the evolution for $\tau>0$ is subject to a fingering instability.

More generally, if $F=F(t)$ is time-dependent, then (\ref{force_balance}a) still holds, giving
 \be
 \frac{1}{h^4}=\left(\frac{8\pi\,}{3\mu\,V^2}\right)\int_{0}^{t} F(t')\tn{d}t' \,+\frac{1}{h_{0}^4}.
 \ee
For prescribed $F(t)$, e.g.  $F(t)\sim t^{\,\alpha}$,  it is then a trivial matter to obtain results analogous to (\ref{hoft}); we shall not pursue these possibilities here.

\section{Effect of a trapped air bubble}\label{air_bubble}
Suppose now that a bubble of air is trapped so that the viscous liquid now occupies the annular region $b(t)<r<a(t)$.  The volume of the bubble is $V_{b}=\pi b^{2} h$, and treating the air as incompressible (see \S \ref{Boyle} below), this is constant, so that
\be\label{bubble_radius_rate}
\frac{\tn{d}b}{\tn{d}t}=-\frac{b}{2h}\frac{\tn{d}h}{\tn{d}t}\,.
\ee
  The volume of the viscous liquid is now $V=\pi \left(a^{2}-b^{2}\right)h$, and this is also constant. This leads to only minor changes in the above analysis. Equation (\ref{barur}) obviously remains unchanged. Also the flux balance 
  $ 2\pi r/h= -\pi r^{2} \bar{u}(r,t)\tn{d}h/\tn{d}t$ still holds, since the air as well as the viscous liquid is assumed incompressible. Hence (\ref{p_of_r}) remains valid also and the pressure is still given by (\ref{pressure0})  for  $b<r<a$.  The pressure $p_{b}(t)$ in the air bubble is then given by
  \be\label{p_bubble_2}
  p_{b}(t)=p(b,t)+\gamma/h = p_{a}-\frac{3\mu}{h^3}\frac{\tn{d}h}{\tn{d}t}\left(a^2 -b^2\right)\,.
  \ee

  The force $F=F_{p}$ is now given by
  \be
  F=\int_{b}^{\,a}\left(p(r,t)-p_{a}\right)\,2\pi r\, \tn{d}r +(p_{b}-p_{a})\,\pi b^2 \,.
  \ee
Using (\ref{pressure0}) and (\ref{p_bubble_2}), this evaluates to
\be
F=-\frac{3\pi\mu}{2h^3}\frac{\tn{d}h}{\tn{d}t}\left(a^2 -b^2\right)\left(a^2 +b^2\right) -\frac{\pi\gamma}{h^2}\left(a^2 -b^2\right)
= \frac{3\mu V(V+2V_{b})}{8\pi}\frac{\tn{d}}{\tn{d}t}\left(\frac{1}{h^4}\right) -\frac{\gamma\, V}{h^2}\,,
\ee
which reduces as expected to (\ref{force_balance}) when $V_{b}=0$.

Thus again, with  $t_{0}$ now given by
\be\label{time-scale_2}
t_{0}= \frac{3\mu}{8\pi F h_{0}^4}V(V+2V_{b})\,,
\ee
we arrive at the following trivial modifications of (\ref{hoft}) and (\ref{implicit_X_2}), if surface tension is neglected:
\bea\label{hoft_2}
 h(\tau)&=& h_{0}(1\!+\! \tau)^{-1/4}\,, \, a(\tau)=a_{0}(1\!+\! \tau)^{1/8}\,,\,  b(\tau)=b_{0}(1\!+\! \tau)^{1/8}\,,\,\,\tn{if}\,\,F>0\,\tn{with}\,\tau=t/t_{0}\,,\\ \label{hoft_3}
 h(\tau)&=& h_{0}(1\!-\! \tau)^{-1/4}\,,\,  a(\tau)=a_{0}(1\!-\!\tau)^{1/8}\,,\,  b(\tau)=b_{0}(1\!-\!\tau)^{1/8}\,,\,\,\tn{if}\, F<0\, \tn{with}\,\tau= t/|t_{0}|\,.
\eea
Thus the presence of the bubble merely affects the time-scale of this evolution, and the one-eighth power law still applies [now to both $a(\tau)$ and $b(\tau)$].  We shall find however that the stability of these evolving states is seriously affected by the presence of the bubble.
\subsection{Neglect of compressibility of air in the bubble}\label{Boyle}
Note that, from (\ref{p_bubble_2}), when $\tn{d}h/\tn{d}t<0$, $p_{b}(t)>p_{a}$, i.e.~the pressure in the bubble is greater than atmospheric (this is essentially because $\bar{u}_{r} >0$, so that, according to (\ref{barur}), $p(r,t)$ is monotonic decreasing in $r$); the air in the bubble is therefore slightly compressed. This effect is however small in the experiments described in \S\ref{Sec_exp} below; for, using (\ref{time-scale_2}) and (\ref{hoft_2}), (\ref{p_bubble_2}) becomes
 \be\label{p_bubble_3}
 p_{b}(\tau)=p_{a} +\frac{2F\,h_{0}}{\left(V+V_{b}\right)(1+\tau)^{9/4}}\,.
 \ee
With the estimates $F\sim 10$ kg\,m\,s$^{-2}$, $V+V_{b}\sim 10^{-5}$m$^{3}$ and $h_{0}\sim 2\times 10^{-3}$m, we find that 
$2Fh_{0}/(V+V_{b}) \sim 4\times 10^{3}$ kg m$^{-1}$s$^{-2}$,  small compared with atmospheric pressure 
$p_{a}\sim 10^{5}$ kg m$^{-1}$s$^{-2}$. Thus at time $\tau$, $[p_{b}(\tau)-p_{a}]/p_{a}\sim 0.04 (1+\tau)^{-9/4}$ and Boyle's Law implies a correspondingly small reduction in the volume of the bubble due to compression.  The compressibility of the air in the bubble is therefore negligible in the conditions of our experiment, although other circumstances can be imagined when the effect might become significant.

\section{Stability of the basic state}\label{Sec_stability}
In this section, we suppose $F>0$, and for simplicity we neglect the effect of surface tension.
 We  suppose first that the viscous liquid covers a simply-connected domain $S\!(t)$ in the $\{x,y\}$-plane, bounded by the curve $C(t)\!:\! r=R(\theta, t)$, where $x=r\cos\theta,y=r\sin\theta$, and  $R(\theta, t)$ is single-valued and periodic in $\theta$, with period $2\pi$.  The area of $S\!(t)$ is
 \be
 A(t)= \iint_{S} \tn{d}x\,\tn{d}y =\int_{0}^{2\pi}\left[\int_{0}^{R(\theta,t)}r\,\tn{d}r\right]\,\tn{d}\theta = \thalf\int_{0}^{2\pi}R^{2}(\theta,t)\,\tn{d}\theta\,,
 \ee
 and the volume of the drop is $V=h(t)A(t)$.  With origin $r=0$ at the centre-of-mass of the drop, 
 \be\label{CG}
  \iint_{S(t)}\{x,y\}\, \tn{d}x\,\tn{d}y =\tthird \int_{0}^{2\pi} R^{3}(\theta, t)\{\cos\theta, \sin\theta\} \,\tn{d}\theta = 0\,.
 \ee
 
 The pressure $p=p(r,\theta)$ still satisfies Reynolds' equation (\ref{Reynolds}) so that
 \be\label{Reynolds_2}
 \nabla^{2}p = \frac{12\,\mu}{h^{3}}\frac{\tn{d}h}{\tn{d}t}\,,\quad p=0 \,\,\tn{on} \,\, C(t)\,.
 \ee 
 If $C(t)$ is known at any instant $t$, this Dirichlet problem may be solved numerically; this provides the basis for a step-by-step determination of the evolution of $C(t)$ from an initial condition $C(0)=C_{0}$, say.

 \subsection{Elliptic drop}\label{Sec_elliptic_drop}
 \begin{figure}
\begin{center}
\includegraphics*[width=0.7\textwidth,  trim=0mm 0mm 0mm 0mm]{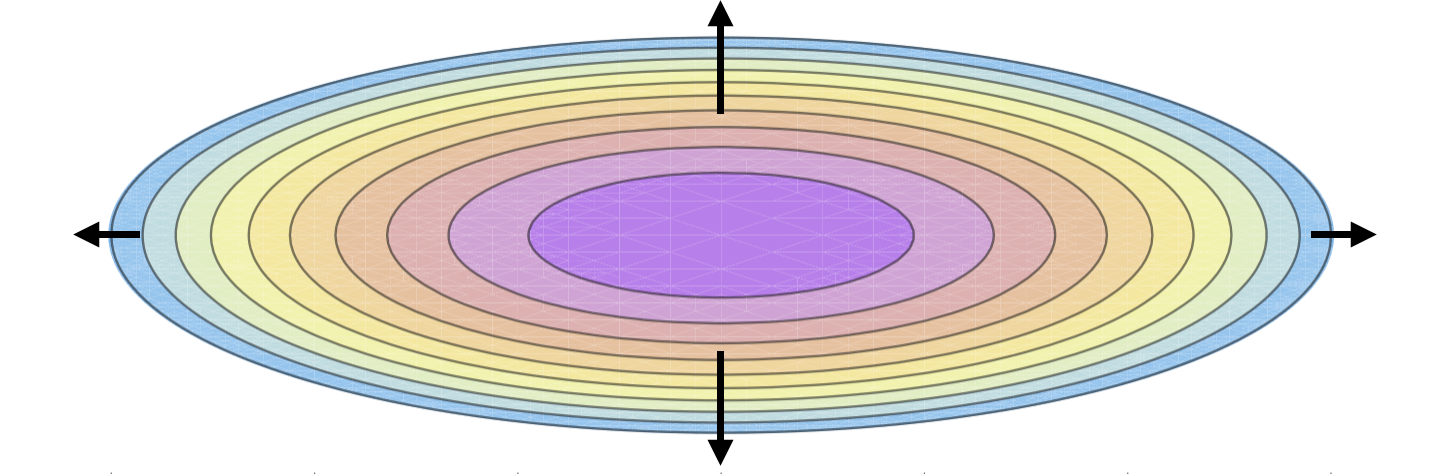} 
\end{center}
\caption{Pressure contours $p=$ const~for the particular  initial condition $a_{0}/b_{0}=3$, as given by (\ref{p_elliptic}); the  magnitude of the pressure gradient at the boundary is indicated by the length of the arrows.} 
  \label{Fig_elliptic_drop}
\end{figure}
 The situation is well illustrated by the case of an  elliptic drop, for which $C(t)$ has the form
 \be
\frac{ x^{2}}{a^{2}(t)} + \frac{ y^{2}}{b^{2}(t)}  =1\,,\quad\tn{with}\quad a(0)=a_{0},\,b(0)=b_{0}, 
\ee
 and where
 \be
 \pi \,a(t)b(t)h(t)= V=\tn{constant} =\pi \,a_{0}b_{0}h_{0}\,.
\ee
We may assume $b_{0}\le a_{0}$. The pressure is given by
\be\label{p_elliptic}
p(x,y,t)-p_{a}=\hat {p}(t)\left(\frac{ x^{2}}{a^{2}(t)} + \frac{ y^{2}}{b^{2}(t)} -1\right),
\ee
so that, using (\ref{Reynolds_2}),
\be
\nabla^{2}p= 2 \hat {p}(t)\left(\frac{ 1}{a^{2}(t)} + \frac{ 1}{b^{2}(t)} \right)=\frac{12\,\mu}{h^{3}}\frac{\tn{d}h}{\tn{d}t}\,.
\ee
Hence
\be\label{p_hat}
\hat{p}(t)=\frac{6\,\mu}{h^{3}}\frac{\tn{d}h}{\tn{d}t} \frac{a^2 b^2}{a^2 +b^2 } 
= \frac{6\,\mu\,V^2}{\pi^2 h^{5}}\frac{\tn{d}h}{\tn{d}t}\frac{1}{ a^2 +b^2 }\,.
\ee
The downward force $F$ applied to the upper plate is now
\be\label{force_ellipse}
F=\iint_{S(t)}\hat {p}(t)\left(\frac{ x^{2}}{a^{2}(t)} + \frac{ y^{2}}{b^{2}(t)} -1\right)\,\tn{d}x\,\tn{d}y=-(1/2) \hat {p}(t)\,\pi a b
=-(V/2)\hat {p}(t)/h(t)\,.
\ee

The pressure contours $p=$ const.~are shown in figure \ref{Fig_elliptic_drop}.  The pressure gradient at the boundary, as shown by the arrows, is greater where these contours are close together;  the rate of expansion in the $y-$direction is therefore greater than  in the $x-$direction.  We can calculate these rates as follows.
 \begin{figure}
\begin{center}
\includegraphics*[width=0.45\textwidth,  trim=0mm 0mm 0mm 0mm]{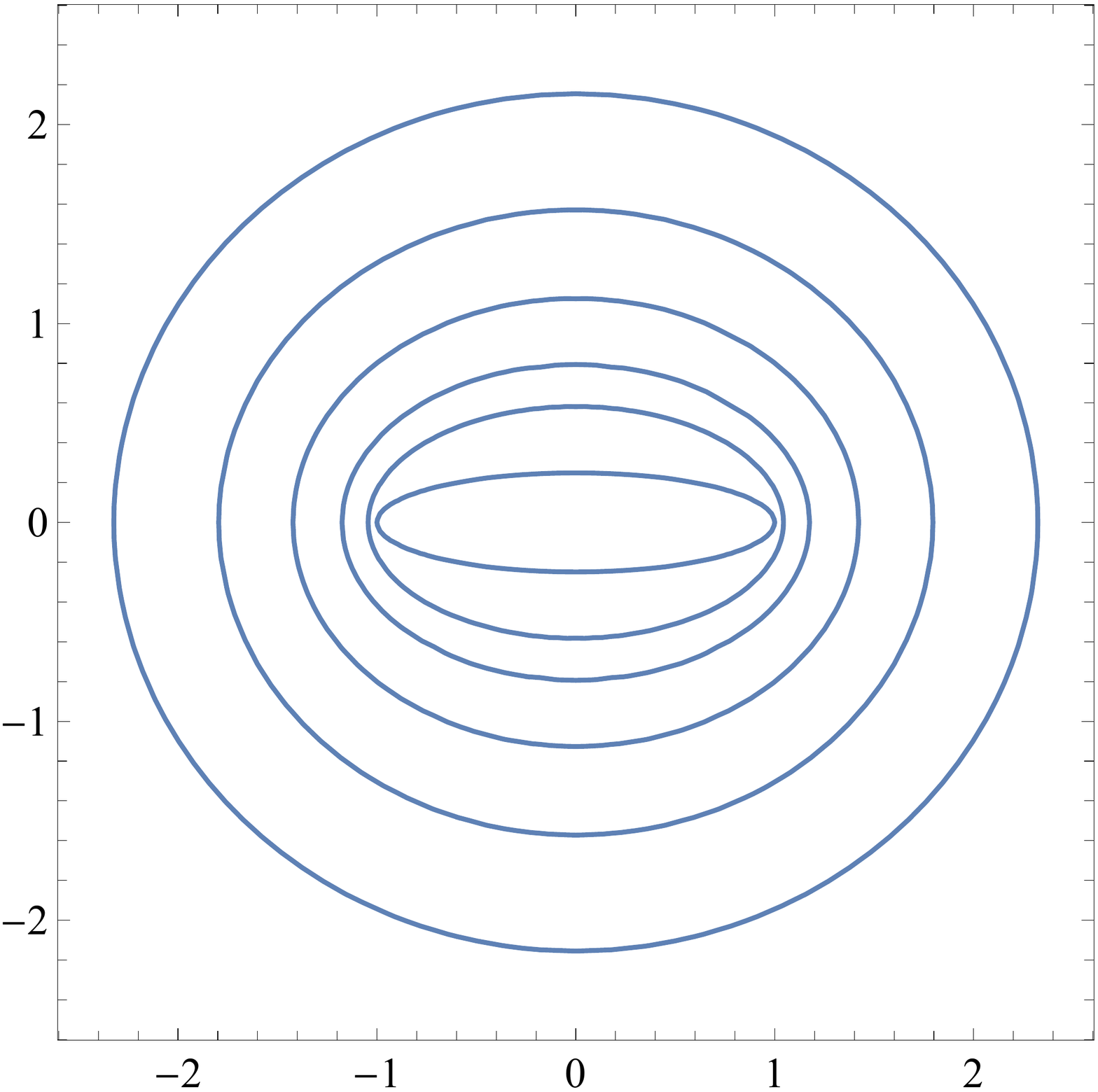} \quad
\includegraphics*[width=0.45\textwidth,  trim=0mm 0mm 0mm 0mm]{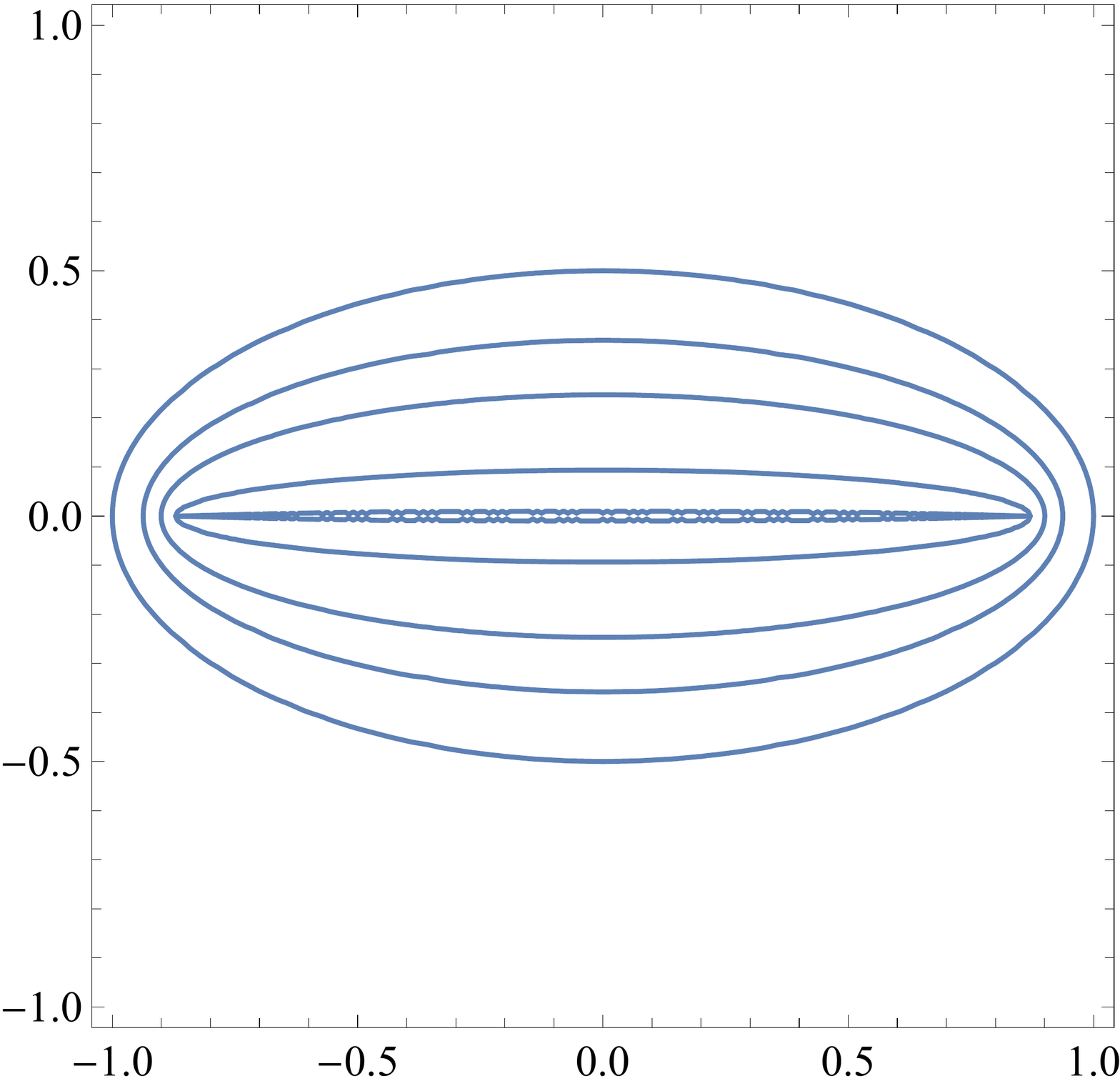} \\
(a)\qquad\qquad\qquad\qquad \qquad\qquad\qquad\qquad\quad (b)
\end{center}
\caption{Evolution of an elliptic drop, with initial condition $a_{0}=1,b_{0}=0.5,\kappa=0.75$; (a) expanding drop when $F > 0$, shown at times $\tau=0,1,10,10^2,10^3$ and $10^4$; the shape becomes gradually more circular in form;  (b) contracting drop when $F < 0$; the minor axis of the ellipse $b(\tau)$ decreases to zero at the singularity time $\tau=0.71672356\dots$; the contracting boundary, shown at times $\tau= 0,0.6,0.7,0.7166,0.71672356$, is subject to smaller-scale instabilities.} 
  \label{Fig_expanding_ellipse}
\end{figure}
The $x$-component of velocity at the point $(a,0)$ is
\be
\frac{\tn{d}a}{\tn{d}t}= \bar{u}(x,0)= -\frac{h^2}{12\mu}\left.\frac{\partial p}{\partial x}\right|_{x=a,\,y=0}=-\frac{h^2}{6\mu a}\hat{p}(t)
=-\frac{1}{ a} \frac{V^2}{\pi^2 h^{3}}\frac{\tn{d}h}{\tn{d}t}\frac{1}{ (a^2 +b^2 )}
\ee
and hence
\be\label{a_squared}
\frac{\tn{d}a^2}{\tn{d}t}= \frac{V^2}{\pi^2}\frac{\tn{d}}{\tn{d}t}\left(\frac{1}{h^2}\right)\frac{1}{ \left(a^2 +b^2 \right)}\,.
\ee
Similarly,
\be
\frac{\tn{d}b^2}{\tn{d}t}= \frac{V^2}{\pi^2}\frac{\tn{d}}{\tn{d}t}\left(\frac{1}{h^2}\right)\frac{1}{ \left(a^2 +b^2 \right)}
\ee
and it  follows that $a^2 (t)-b^2 (t)=$ constant $=a_{0}^2-b_{0}^2$.  We then have from (\ref{a_squared})
\be
\frac{\tn{d}}{\tn{d}t}\left[a^4 -\left(a_{0}^2-b_{0}^2\right)a^2\right]= \frac{V^2}{\pi^2}\frac{\tn{d}}{\tn{d}t}\left(\frac{1}{h^2}\right),
\ee
so that, with the initial condition $a(0)=a_{0}$, $a^2$ satisfies the quadratic equation
\be
a^4 -(a_{0}^2-b_{0}^2)a^2=V^2/\pi^2h^2\,,
\ee
with relevant root
\be\label{a_squared_2}
a^2(t)= \thalf \left(a_{0}^2-b_{0}^2\right) + \thalf \left[ \left(a_{0}^2 - b_{0}^2\right)^{2}+4V^2/\pi^2 h^2\right]^{1/2}\,.
\ee
The corresponding expression for $b^2(t)$ is
\be\label{b_squared}
b^2 (t)=a^2 (t)-\left(a_{0}^2-b_{0}^2\right)= -\thalf \left(a_{0}^2-b_{0}^2\right) 
+ \thalf \left[ \left(a_{0}^2-b_{0}^2\right)^{2}+4V^2/\pi^2h^2\right]^{1/2}\,,
\ee
so that
\be
a^2(t) + b^2 (t)= \left[ \left(a_{0}^2-b_{0}^2\right)^{2}+4V^2/\pi^2h^2\right]^{1/2}\,.
\ee
From (\ref{p_hat}) we then have
\be
\hat{p}(t)=
- \frac{6\,\mu\,V^2}{\pi h^{4}}\frac{\tn{d}h}{\tn{d}t} \left[ \left(a_{0}^2-b_{0}^2\right)^{2}\pi^2h^2+4V^2\right]^{-1/2}\,.
\ee
and  from (\ref{force_ellipse}),
\be\label{force_ellipse}
F=\frac{3\,\mu\,V^2}{8\pi}\frac{\tn{d}}{\tn{d}t}
 \left(\frac{1}{h^4}\right) \left[ 1+\left\{\left(a_{0}^2-b_{0}^2\right)\pi h/2V\right\}^2\right]^{-1/2}\,,
\ee
in agreement with (\ref{force_balance}) when  $a_{0}=b_{0}$.  

In dimensionless form, with $\tau=t/t_{0}$  [where $t_{0}$ is still defined by (\ref{t_zero})] and $\hat{h}(\tau)=h(t)/h_{0}$, (\ref{force_ellipse}) becomes
\be
\frac{\tn{d}}{\tn{d}\tau} \left(\frac{1}{\hat{h}^4}\right)= \left[ 1+\kappa^{2} \hat{h}^2\right]^{1/2}\,,
\ee
where
\be
\kappa=\left(a_{0}^2-b_{0}^2\right) /2 a_{0}b_{0}\,.
\ee
Here, $\kappa \ge 0$ is  a measure of the initial eccentricity of the ellipse.
With this notation, (\ref{a_squared_2}) and  (\ref{b_squared}) become
\be
a^2(\tau)/a_{0}b_{0}= \kappa+ \left[\kappa^{2}+ \hat{h}^{-2}(\tau)\right]^{1/2}\,,\quad b^2(\tau)/a_{0}b_{0}= -\kappa + \left[\kappa^{2}+ \hat{h}^{-2}(\tau)\right]^{1/2}\,,
\ee

The evolution is shown in figure \ref{Fig_expanding_ellipse} for initial conditions $a_{0}=1,b_{0}=0.5$ (so $\kappa=0.75$). In (a), $F>0$, and the drop expands, becoming gradually more circular in form.  In (b),  $F<0$, and the drop contracts with increasing ellipticity [as $\hat{h}(\tau)$ increases], becoming singular as $\tau\rightarrow \tau_c = 0.71672\dots\,$, when  $b(\tau)\rightarrow 0$ and $a(t)\rightarrow \surd {3}/2 =0.8660\dots\,$. However, this behaviour persists only for so long as $b(\tau)\gtrsim h_{0} \,\hat{h}(\tau)$, and it is moreover subject to a `fingering instability'.  Surface tension must also become important where the curvature of the boundary becomes large.  These effects will be considered  in \S \ref{Sec_fingering_instability} below.

 \subsection{Linearised stability analysis}\label{linearised_stab}
We here consider a circular drop with perturbed boundary
 \be\label{R_zero}
 R(\theta, 0) =a_{0} +\epsilon \,\alpha_{0} \cos n\theta
 \ee
at the initial instant, where $n$ is a positive integer.  When $n=1$, the circle is merely displaced in the $x$-direction without distortion;  we may therefore suppose that $n \ge 2$, so that (\ref{CG}) is automatically satisfied. (When $n=2$, the circle is perturbed to an ellipse.)  With the  choice (\ref{R_zero}),
 \be
 A(0)= \pi a_{0}^{2} +\thalf\pi\, \epsilon^{2}  \alpha_{0}^2.
 \ee
 
  When $0<\epsilon \ll 1$, the problem may be linearised.  Then for $t>0$,
 \be
  R(\theta,t) =a(t) +\epsilon\, \alpha(t)  \cos n\theta\,,
 \ee
 where $a(t)$ is given by (\ref{hoft}) and $A(t)=\pi a^{2}(t) +\tn{O}(\epsilon^{2})$. Thus, let
 \be
 p(r,\theta,t) =p_{0}(r,t) +\epsilon\, p_{1}(r,t) \cos n\theta\,,
 \ee
where  $p_{0}(r,t)$ is given by (\ref{pressure0}).  At this level of approximation, $h(t)$ is still given by (\ref{hoft}), and since
$ \nabla^{2}p_{0} =12\,\mu\, h^{-3}\tn{d}h/\tn{d}t$, (\ref{Reynolds}) gives
\be
\frac{1}{r}\frac{\partial}{\partial r} r \frac{\partial p_{1}}{\partial r} -\frac{n^2}{r^2}p_{1} =0\,.
\ee
The solution, finite at $r=0$, is
\be
p_{1}(r,t) =k(t)r^{n}\,,
\ee 
so that now
\be
p(r,\theta,t)= p_{0}(r,t) +\epsilon\, k(t)r^{n} \cos n\theta\,.
\ee
The pressure boundary condition is now $p(r,\theta,t)-p_{a}=0$ on $r=a(t)+\epsilon \,\alpha \cos n\theta$, and, noting that
\be
 p_{0}(a(t)+\epsilon\, \alpha(t)\cos n\theta,t)-p_{a} =  p_{0}(a)-p_{a} +\epsilon\, \alpha(t)\cos n\theta \left. \frac{\partial p_{0}}{\partial r}\right|_{r=a} = \epsilon\, a(t)\,\alpha(t)\cos n\theta\frac{6\mu }{h^3}\frac{\tn{d}h}{\tn{d}t}
\ee
and that
\be
\epsilon\, p_{1}(a(t)+\epsilon\, \alpha(t)\cos n\theta,t) =\epsilon\, k(t)a^{n} + \tn{O}(\epsilon^2)\,,
\ee
this boundary condition (at order $\epsilon$) determines $k(t)$, leading finally to
\be
p(r, \theta,t)-p_{a}=\left[(r^2 -a^2) -2 \epsilon \,a \,(r/a)^{n}\,\alpha(t)\cos n\theta\right]\frac{3\mu}{h^3}\frac{\tn{d}h}{\tn{d}t} +\tn{O}(\epsilon^2)\,,
\ee
and so
\be
\frac{\partial p}{\partial r} =\left [r- \epsilon\,n\, \alpha(t) (r/a)^{n-1}\cos n\theta\right]\frac{6\mu}{h^3}\frac{\tn{d}h}{\tn{d}t}+\tn{O}(\epsilon^2)\,.
\ee
On $r=R(\theta,t)=a+\epsilon\, \alpha \cos n\theta$, this therefore gives
\be
\frac{\partial p}{\partial r} = [a-\epsilon\, (n-1) \,\alpha \cos n\theta]\frac{6\mu}{h^3}\frac{\tn{d}h}{\tn{d}t} +\tn{O}(\epsilon^2)\,.
\ee

Now the curve $C(t)$ moves according to the equation 
\be\label{cevol}
\frac{\partial R}{\partial t} = \frac{\tn{d}a}{\tn{d}t}+ \epsilon \, \frac{\tn{d}\alpha}{\tn{d}t}\cos n\theta = \bar u_{r} (R, t) +\tn{O}(\epsilon^2)\,,
\ee
and $ \bar u_{r}$  still satisfies (\ref{barur}), i.e.
\be
 \bar u_{r} = -\frac{h^{2}}{12\mu}\,\left. \frac{ \partial p}{\partial r}\right|_{r=a(t)+\epsilon\, \alpha(t) \cos n\theta}\,.
\ee
At leading order, (\ref{cevol}) merely confirms that $a^{2}(t)\,h(t)=$ const.  At order $\epsilon$, with $h(t)$ given by (\ref{hoft}), we obtain the equation for $\alpha(t)$,
\be\label{db_by_dt}
\frac{\tn{d}\alpha}{\tn{d}t}= \frac{(n-1)}{2h}\frac{\tn{d}h}{\tn{d}t}\,\alpha = -\frac{(n-1)\,\alpha}{8(t+t_{0})}\,,
\ee
which integrates to give
\be
\alpha(t)=\alpha_{0}(1+t/t_{0})^{-(n-1)/8}.
\ee
Since $n\!>\!1$, this ensures that, when $F\!>\!0$ and so $t_{0}\!>\!0$, the perturbation decays to zero as $t\rightarrow\infty$; and in fact the higher the value of $n$, the more rapid is this decay.
\begin{figure}
\begin{center}
\vskip -20mm
\includegraphics*[width=0.45\textwidth,  trim=10mm 80mm 10mm 0mm]{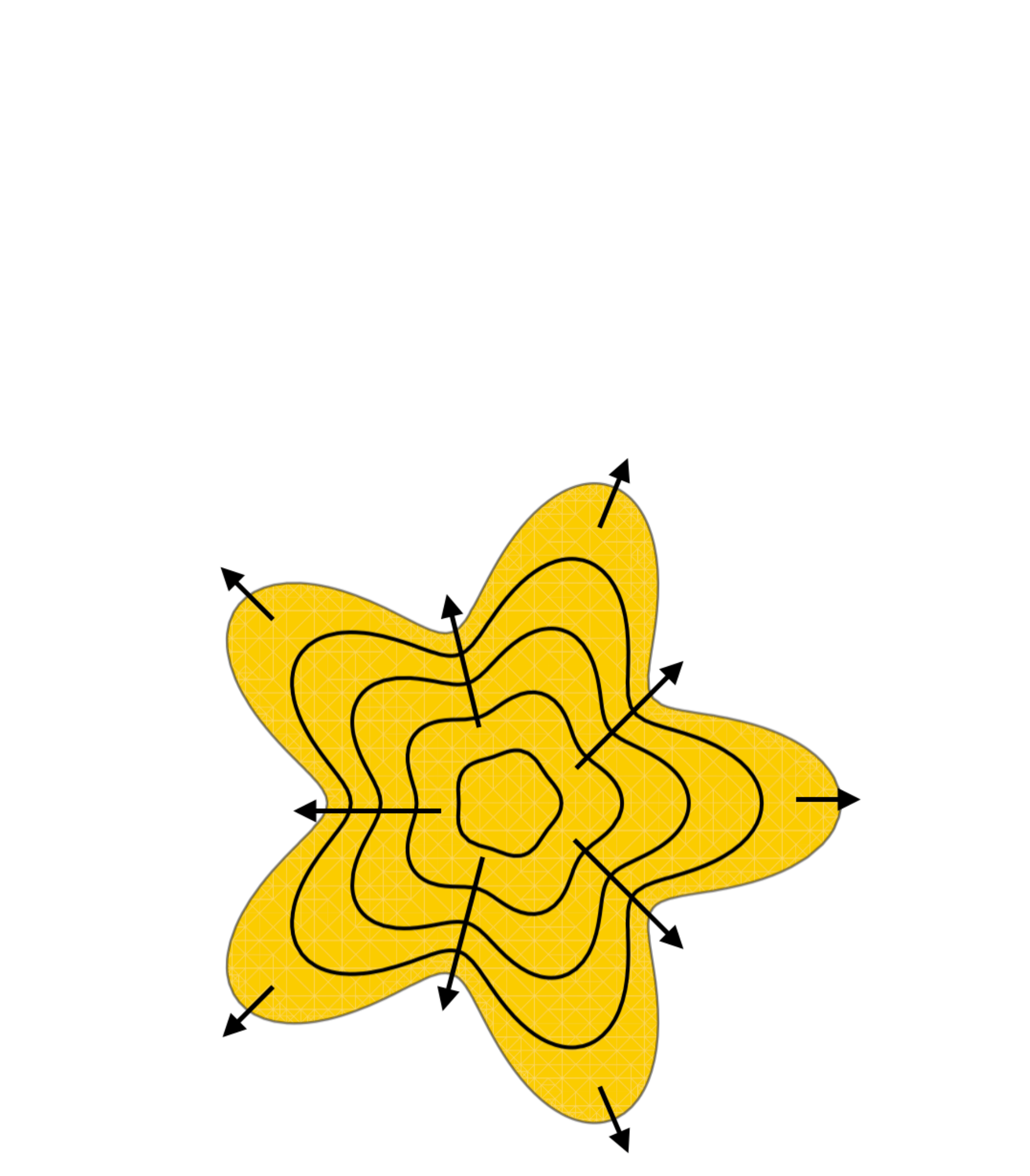} \quad
\includegraphics*[width=0.45\textwidth,  trim=10mm 80mm 10mm 0mm]{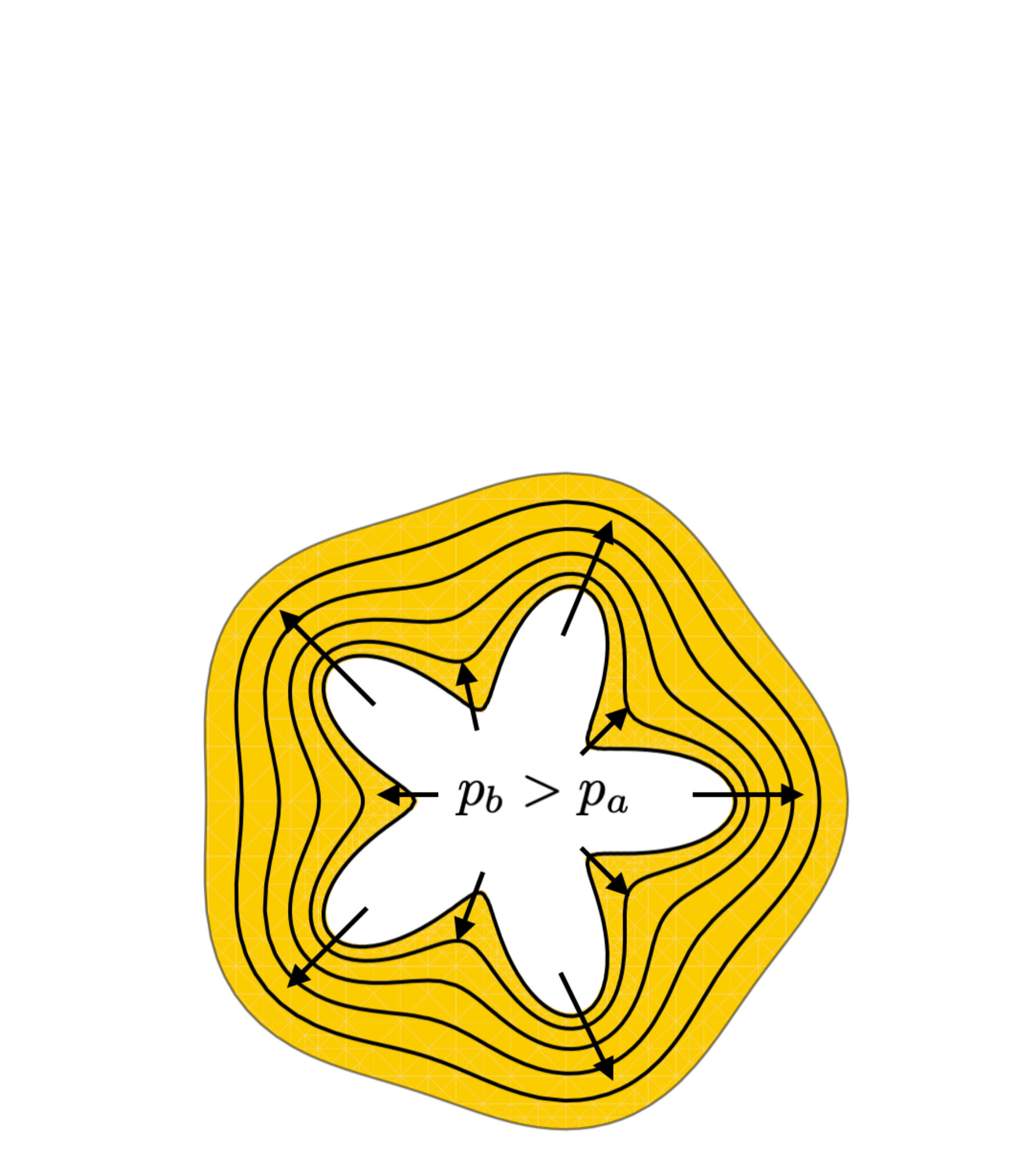} 
 \end{center}
\quad\qquad \qquad\qquad \qquad\quad (a) \qquad\qquad\qquad\qquad\qquad\qquad \qquad\qquad\qquad\qquad (b)
\vskip 2mm
\caption{(a) Sketch indicating the stabilising mechanism for a squeezed drop; the drop  expands more rapidly where the isobars $p=$ const.~are closer together, thus tending to restore a circular shape;  (b) sketch indicating the destabilising mechanism at the boundary of an air bubble, in which the pressure $p_{b}$ is greater than atmospheric pressure $p_{a}$; the expansion is again more rapid where the isobars are closer together, leading now to growth of the perturbation amplitude. }  
 \label{Sketch_ab}
\end{figure}

The stability can be best understood from consideration of the sketch of figure \ref{Sketch_ab}(a) which shows the particular situation when $n=5$ (a five-fold hypotrochoid).  Just as for the case of an elliptic drop  (figure \ref{Fig_elliptic_drop}), the restoring force is greater where the isobars are closer, thus tending to restore a circular shape.  It is here that our conclusion differs from that of \cite{S2015}, as mentioned in the introduction.

\section{Fingering instability when $F\!<0$}\label{Sec_fingering_instability}
\begin{figure}
\begin{center}
\includegraphics*[width=0.8\textwidth,  trim=0mm 50mm 0mm 0mm]{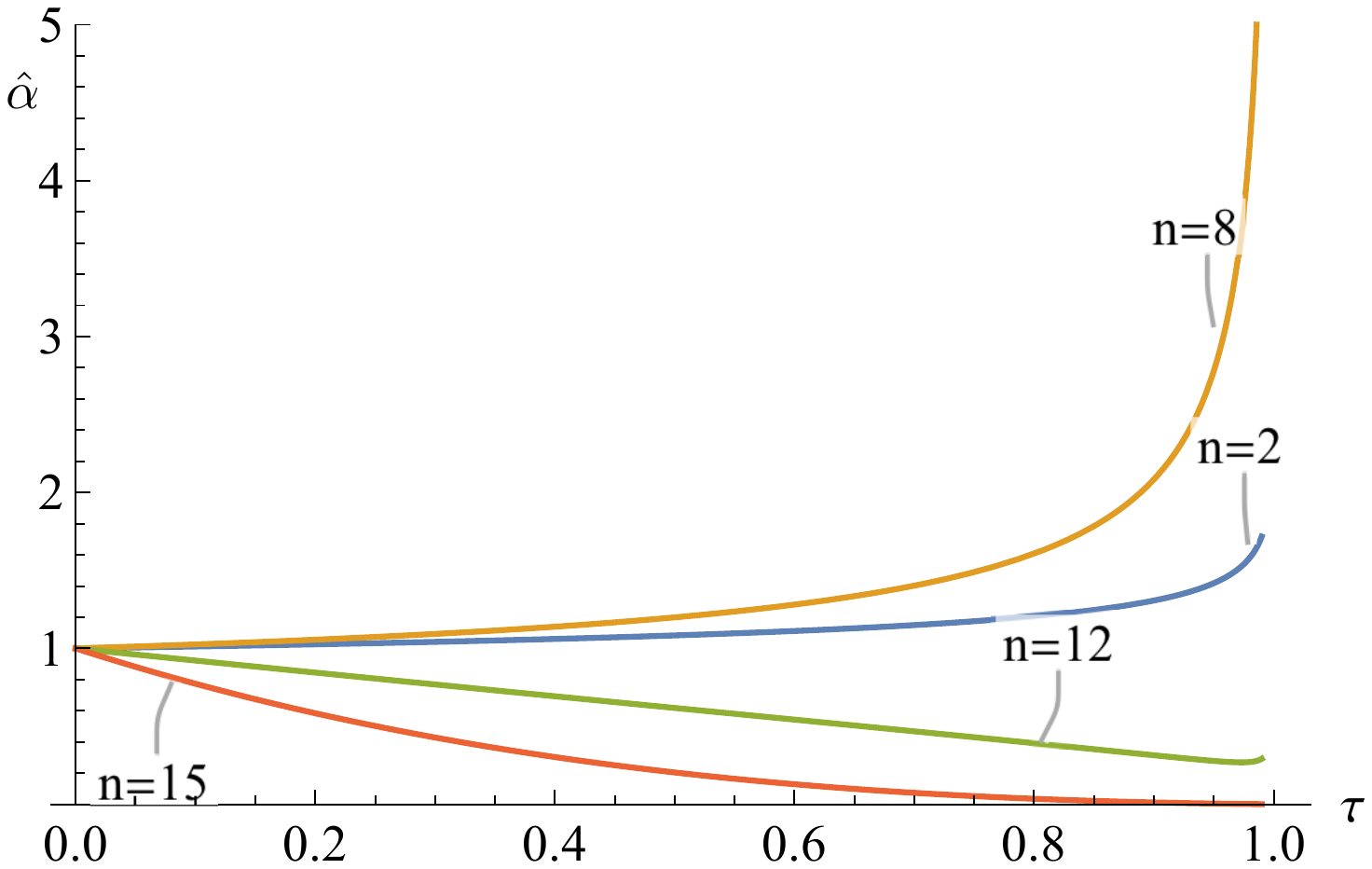} 
\end{center}
\vskip -60mm
\caption{Plot of $\hat{\alpha}(\tau)=\alpha(\tau)/\alpha_{0}$ as a function of $\tau=t/|t_{0}|$, for $\sigma\, |t_{0}|=0.01$ and $n=2, 8,12,15$.} 
  \label{fing_inst}
\end{figure}
When $F\!<0$, the basic state is given by (\ref{implicit_X_2}), so that, for $0<t \ll |t_{0}|$,
\be
\frac{1}{h}\frac{\tn{d}h}{\tn{d}t} = \frac{1}{4(|t_{0}|-t)}\,,
\ee
and (\ref{db_by_dt}) becomes
\be\label{db_by_dt_2}
\frac{\tn{d}\alpha}{\tn{d}t}= \frac{(n-1)\,\alpha}{8(|t_{0}|-t)}\,,
\ee
which integrates to give
\be
\alpha(t)=\alpha_{0} (1-t/|t_{0}|)^{-(n-1)/8}\,.
\ee
This shows the beginning of a `fingering instability' of a type first identified by \cite{S58}, resulting here from the radial contraction of the region occupied by the  viscous liquid and the resulting suction of air into this region.  The instability is stronger as $n$ increases, apparently without limit as $n\rightarrow \infty$.  In reality, surface tension must play a role in determining the most unstable mode, an effect that may be simply treated as follows.

Let $\gamma$ be the surface tension acting at the boundary $r=R(\theta, t) =a+\epsilon\,\alpha \cos n\theta$. As we have seen in \S \ref{basic_state}, this leads to a jump in pressure $\gamma/h(t)$ across the air/liquid interface; this jump, independent of $\theta$, is irrelevant for the stability of the circular shape and can be ignored here.   There is also however a contribution to the pressure jump from the variation of curvature around the drop boundary.  We assume that $\gamma$ is relatively weak, 
so that surface tension is significant only for modes $\sim \!\cos n\theta$ for which $n\gg 1$.  Then the pressure condition $p(r,t)-p_{a}=0$ on $r=R$ should be replaced by
\be
p(r,t)-p_{a} \approx -\gamma \frac{1}{a^2} \frac{\partial^2 R}{\partial \theta^{2}} = \gamma n^{2}\frac{\alpha}{a^2}\cos n\theta \quad \tn{on}\quad r=a+\epsilon\, \alpha \cos n\theta\,.
\ee
This modified boundary condition leads to a corresponding modification in (\ref{db_by_dt_2}), which now becomes, after some simplification,
\be
\frac{\tn{d}\alpha}{\tn{d}t}= \frac{(n-1)\,\alpha}{8|t_{0}|(1-t/|t_{0}|)} -\frac{\sigma\, n^{3}\alpha}{8(1-t/|t_{0}|)^{7/8}} \quad\tn{where}\quad \sigma= \left(\frac{2\gamma}{3\mu}\right)\frac{ h_{0}^{2}}{a_{0}^{3}}\,.
\ee
With $\tau=t/|t_{0}|$, this integrates to give
\be  \label{b_of_t}
\alpha(\tau)= \alpha_{0} (1-\tau)^{-(n-1)/8}   \exp {\left[-n^{3}\sigma\, |t_{0}| \left\{1-(1-\tau)^{1/8}\right\}\right]}\,.
\ee

By way of example, the function $\alpha(\tau)/\alpha_{0}$ is shown in figure \ref{fing_inst}, for the parameter values $\sigma\, |t_{0}|=0.01$
and $n=2,8,12$ and $15$, and for the interval $0 <\tau <0.99$.  The modes become increasingly unstable as $n$ increases from 2 to 8, but by $n=12$, the stabilising effect of surface tension is evident, and the modes are rapidly damped for $n\gtrsim 15$. More generally, surface tension has a strong stabilising effect for modes for which $n\gtrsim (\sigma\, |t_{0}|)^{-1/3}$.

\section{Destabilising effect  of trapped air bubble}\label{Sec_destabilising_bubble}
We here follow the procedure described in \S\ref{linearised_stab} above, to examine the stability of the evolving state with a trapped air bubble, as described in \S\ref{air_bubble}. For simplicity in this section, we ignore surface tension and assume first that $F>0$ so that 
$\tn{d}h/\tn{d}t<0$. In the undisturbed state, the viscous liquid is contained in the region $a(t)>r>b(t)$.  We suppose that the liquid boundaries are perturbed to
\be
r= R_{1}(\theta, t)= a(t) +\epsilon \,\alpha(t) \cos{n\theta}\,,\quad 
r= R_{2}(\theta, t)= b(t) +\epsilon\, \beta(t) \cos{n\theta}\,,
\ee
and  linearise in $\epsilon$.  The pressure field in the region $ R_{1}(\theta, t)>r>R_{2}(\theta, t)$ still takes the form
\be
p(r,\theta,t)=p_{a}+p_{0}(r,t) +\epsilon p_{1}(r,\theta,t)\,,
\ee
where $\nabla^{2}  p_{1}=0$, but now 
\be
p_{1}(r,\theta,t)=- \frac{6\mu}{h^3}\frac{\tn{d}h}{\tn{d}t}\,\left [\hat{k}(t)r^{n}+\hat{m}(t)r^{-n}\right]\cos{n\theta}\,,
\ee
where $\hat{k}(t)$ and  $\hat{m}(t)$ are to be determined.
As before, the condition $p(r,\theta,t)=p_{a}$ on $r=R_{1}(\theta, t)$ leads, at order $\epsilon$, to  
\be\label{k_of_t}
\hat{k}(t)\, a^n(t) +\hat{m}(t)\,a^{-n}(t) = \alpha(t)\,a(t)
\ee
and similarly the condition $p(r,\theta,t)-p_{a}=p_{b}$ on $r=R_{2}(\theta, t)$ leads, at order $\epsilon$, to
\be\label{m_of_t}
\hat{k}(t)\, b^n(t) +\hat{m}(t)\,b^{-n}(t) = \beta(t)\,b(t)\,.
\ee
Solving (\ref{k_of_t}) and (\ref{m_of_t}) for $\hat{k}(t)$ and $\hat{m}(t)$,  we obtain
\be
\hat{k}(t)=\frac{\alpha\, a^{n+1}-\beta\,b^{n+1}}{a^{2n}-b^{2n}}\qquad\tn{and}\quad
\hat{m}(t)=\frac{a^{n}b^{n}\left(\beta\, b\,a^{n}- \alpha\,a\,b^{n}\right)}{a^{2n}-b^{2n}}\,.
\ee

We now have
\be
\frac{\partial p}{\partial r}=\frac{6\mu}{h^3}\frac{\tn{d}h}{\tn{d}t}\,r +\epsilon\left[ n\,\hat{k}\, r^{n-1}-n\,\hat{m}\,r^{-n-1}\right]\cos{n\theta}\,,
\ee
so that, on $r=R_{2}= b +\epsilon\, \beta \cos{n\theta}$, 
\be
\left.\frac{\partial p}{\partial r}\right |_{r=R_{2}}=\frac{6\mu}{h^3}\frac{\tn{d}h}{\tn{d}t}\,( b +\epsilon\, \beta \cos{n\theta})+\epsilon\left[ n\,\hat{k}\,b^{n-1}-n\,\hat{m}\,b^{-n-1}\right]\cos{n\theta} +\tn{O}(\epsilon^2)\,.
\ee
Hence, recalling (\ref{barur}), we have
\be
\left.\bar{u}_{r}\right |_{r=R_{2}}=\left.\bar{u}_{r}\right |_{0}-\frac{1}{2h}\frac{\tn{d}h}{\tn{d}t}\,
\left[\beta +n\left(\hat{k}\,b^{n-1} -\hat{m}\,b^{-n-1}\,\right)\right]\epsilon\,  \cos{n\theta}\,,
\ee
where $\left.\bar{u}_{r}\right |_{0}$ is the velocity in the unperturbed state.
At order $\epsilon$,  $\partial R_{2}/\partial t = \left.\bar{u}_{r}\right |_{r=R_{2}}$ now leads to   
\be
\frac{\tn{d}\beta}{\tn{d}t}= -\frac{1}{2h}\frac{\tn{d}h}{\tn{d}t}\,\left\{\beta-n\left[\frac{2 a^{n+1}b^{n-1}\alpha-\left (a^{2n}+b^{2n}\right )\beta}{a^{2n}-b^{2n}}\right]\right\}\,.
\ee
Similarly,  $\partial R_{1}/\partial t = \left.\bar{u}_{r}\right |_{r=R_{1}}$  leads at order $\epsilon$ to
\be
\frac{\tn{d}\alpha}{\tn{d}t}= -\frac{1}{2h}\frac{\tn{d}h}{\tn{d}t}\,\left\{\alpha-n\left[\frac{\left (a^{2n}+b^{2n}\right )\alpha-2 a^{n-1}b^{n+1}\beta}{a^{2n}-b^{2n}}\right]\right\}\,.
\ee
With dimensionless time $\tau=t/t_{0}$, noting from (\ref{hoft_2}) that
\be
 -\frac{1}{2h}\frac{\tn{d}h}{\tn{d}t}=\frac{1}{8(1+\tau)}\quad\tn{and}\quad \frac {b(\tau)}{a(\tau)} =\tn{const.}=\eta,\,\tn{say,}\quad \tn{where}\,\, 0<\eta<1\,, 
\ee
these equations may be written in the form
\be\label{eqns_alpha_beta}
\frac{\tn{d}\alpha}{\tn{d}\tau} =\frac{1}{8(1+\tau)}(A\,\alpha +B\,\beta)\,,\quad \frac{\tn{d}\beta}{\tn{d}\tau} =\frac{1}{8(1+\tau)}(C\,\alpha +D\,\beta)\,,
\ee
where
\be\label{ABCD}
A=1-n\left(\frac{1+\eta^{2n}}{1-\eta^{2n}}\right)\,,\quad B=\frac{2n\,\eta^{n+1}}{1-\eta^{2n}}\,,
\quad C=-\frac{2n\,\eta^{n-1}}{1-\eta^{2n}}\,,\quad D=1+n\left(\frac{1+\eta^{2n}}{1-\eta^{2n}}\right)\,.
\ee
Note that
\be\label{ABCD_2}
A+D=2,\quad \tn{and}\quad AD-BC =1-n^2\,.
\ee

The equations (\ref{eqns_alpha_beta}) admit solutions of the form $(\alpha,\beta)=\left(\hat{\alpha},\hat{\beta}\right) \,(1+\tau)^{\sigma}$, provided
\be
8\sigma\,\hat{\alpha}=A\,\hat{\alpha}+B\,\hat{\beta}\,,\quad  8\sigma \,\hat{\beta}=C\,\hat{\alpha}+D\,\hat{\beta}\,.
\ee
The determinant condition for a non-trivial solution reduces to
\be
8\sigma=\thalf (A+D)\pm \sqrt{1-(AD-BC)}=1\pm n\,,
\ee
using the results (\ref{ABCD_2}).  The modes for which $\sigma=\sigma _1=(1+n)/8$ and $\sigma=\sigma _2=(1-n)/8$ are respectively unstable and stable, and the corresponding ratios $\hat{\alpha}/\hat{\beta}$ for these modes simplify to
\be \label{amplitude_ratios}
\left[\hat{\alpha}/\hat{\beta}\right]_{1}=\eta^{n+1}\quad \tn{and}\quad \left[\hat{\alpha}/\hat{\beta}\right]_{2}=\eta^{1-n}.
\ee
For example, if $\eta=\thalf$ and $n=5$, then $\left[\hat{\alpha}/\hat{\beta}\right]_{1}=1/64$ and  $\left[\hat{\alpha}/\hat{\beta}\right]_{2}=16$.  Thus, the unstable mode has much larger amplitude at the bubble boundary $r=b$, whereas the stable mode has much larger amplitude at the outer  boundary $r=a$;  we may think of the instability as being `driven' from the bubble boundary.  These properties become stronger as $\eta$ decreases and/or $n$ increases.

\begin{figure}
\begin{center}
\includegraphics*[width=0.5\textwidth,  trim=0mm 0mm 0mm 0mm]{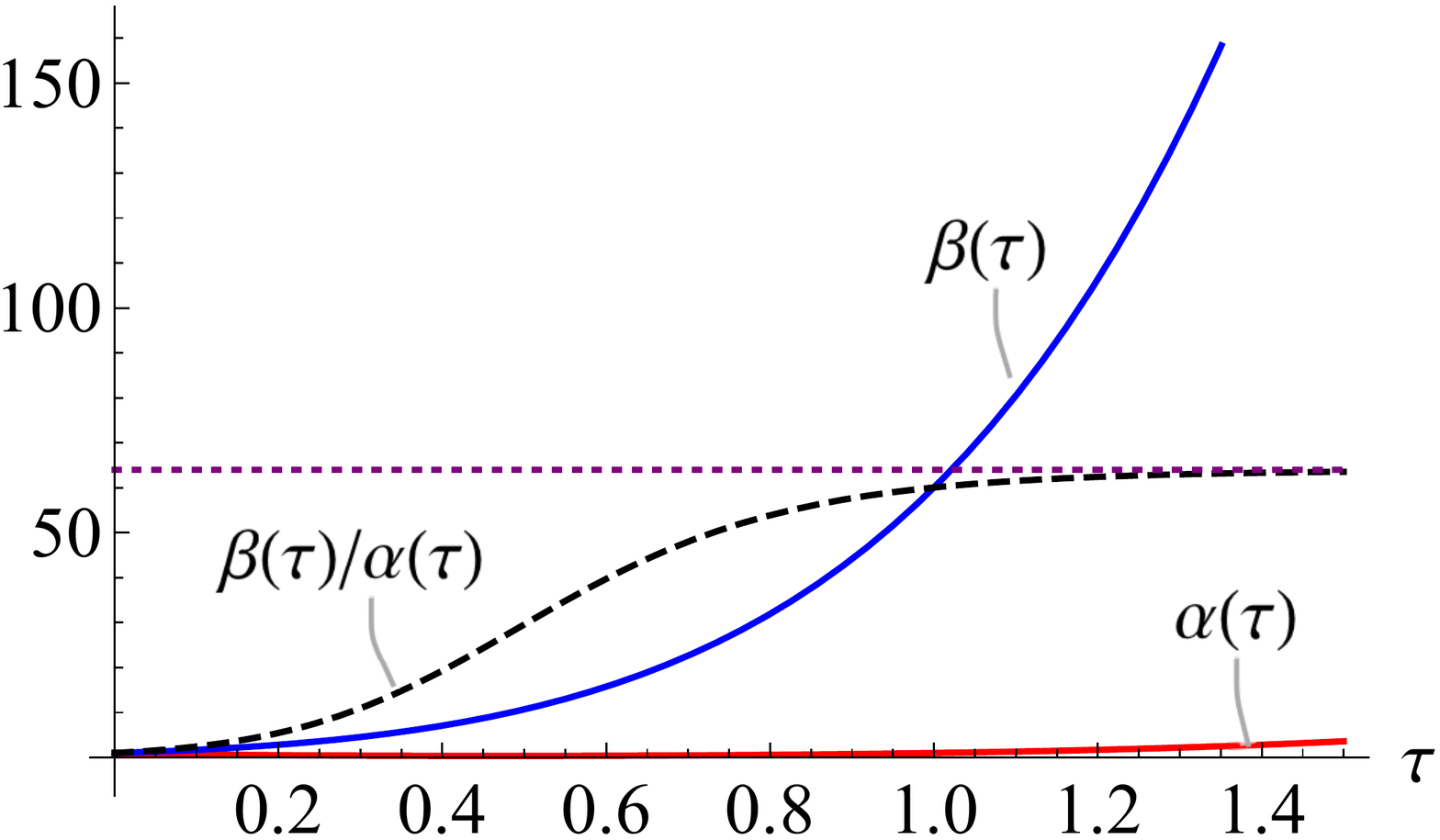} 
 \end{center} 
 \vskip -20mm
 \caption{Solution of equations (\ref{eqns_alpha_beta}) for $\alpha (\tau)$ and $\beta (\tau)$ with initial conditions $\alpha(0)=\beta(0)=1$ and with $\eta=1/2,\,n=5$, as in the sketch of figure \ref{Sketch_ab}(b);  both $\alpha (\tau)$ and $\beta (\tau)$ ultimately increase at the same rate, the ratio $\beta (\tau)/\alpha(\tau)$ tending to a constant for large $\tau$. This behaviour is typical. } 
\label{Growth_instability}
\end{figure}

The general solution corresponding to arbitrary initial conditions $\alpha(0)=\alpha_{0}, \beta(0)=\beta_{0}$ is of course a superposition of these modal solutions, but the unstable mode rapidly dominates. The mechanism of this instability is indicated in the sketch of figure \ref{Sketch_ab}(b): the air bubble is expanding in the mean, but the expansion is more rapid where the pressure contours (isobars) are closer together; this obviously leads to increase of the perturbation amplitude.  

Figure \ref{Growth_instability} shows the behaviour when $\alpha(0)=\beta(0)=1, \eta=b/a=1/2$ and $n=5$. Here, the unstable mode, for which $\beta (\tau)/\alpha(\tau)=64$, dominates for $\tau\gtrsim 1.4$.  The growth rate of this type of instability increases with $n$, and is controlled for large $n$ by surface tension, just as in \S\ref{Sec_fingering_instability}; we need not labour the details here.
\subsection{The situation when $F<0$}\label{situation_bubble_F<0}
When $F<0$, with the basic state given by (\ref{hoft_3}) there are still two perturbation modes, one stable and one unstable, but the unstable mode is now `driven' by the fingering instability at the outer boundary $r=a(t)$, where the amplitude is much greater.

\section{Simple experimental demonstrations}\label{Sec_exp}
\subsection{Single expanding drop}
As already mentioned, our `home experiments' were carried out under effectively lockdown conditions.The first experiment was designed to verify the one-eighth power law (\ref{hoft}). For fluid we used Lyle's Black 
Treacle\footnote{\texttt{http://fiches.ranson.be/00006253nl-be.pdf}} (density 1.41 g/ml,\, viscosity $\mu\approx 64.3\, \tn{kg} \tn{ s}^{-1}\tn{m}^{-1} $ at 22$^{o}$C). A drop of volume $V= 5$ ml  was placed on a horizontal glass plate and a second glass plate of mass $M=1.127$ kg was placed on this and allowed to settle, so that the downward force was
 $F=Mg = 11.04  \,{\tn{kg}\, \tn{m}\,\tn{s}^{-2}} $; thus $\mu/F = 5.82\, \tn{m}^{-2}\tn{s}$.  The plates were illuminated from below, and a camera was set to take shots from above at 10-second intervals. the initial time  $t=0$ was set when the drop radius was $a_{0}=30.72$ mm; thus $ h_{0}= V/\pi a_{0}^{2}=1.69$ mm, and $h_{0}/a_{0}=0.055$. From (\ref{t_zero}), the time-scale $t_{0}$ for this run was thus
\be
t_0=  \frac{3\pi^3\,(5.82) (0.03072)^8}{8(5\times 10^{-6})^2} = 2.15\,\tn{s}\,.
\ee
Figure \ref{Fig_expanding_drop} shows photos of the expanding drop at three times $\tau=t/t_{0}=$ 60,\,\,172, and 321.  The background grid allowed measurement of the diameter in each case.  Figure \ref{Fig_one-eighth}(a) shows the growth of the drop radius for $0\le\tau\le 321$.  The measurements agree well with the result (\ref{hoft}) as shown by the red curve; and figure \ref{Fig_one-eighth}(b) shows that the one-eighth power law is remarkably well satisfied over this full time range.

\subsection{Contraction of drop by leverage}
Figure \ref{Fig_Single_drop_fingering} shows the effect of gently levering the plates open at the lower right-hand corner some time after the stage of figure \ref{Fig_expanding_drop}(c) was reached. The fingering instability immediately develops and the fingers grow as the levering is gradually increased. The duration of course depends on the speed at which the angle between the plates is increased, and as the levering was applied by hand, exact repetition was not possible.  This particular sequence was taken over a period of approximately one minute, and is quite typical. The characteristic tip-splitting and side-branching of fingers is evident, and cavitation bubbles can be detected in the very low pressure region in panel (e).  The reason for this cavitation may be understood with reference to the `hinged-plate' problem briefly considered in \S2 of \cite{M64}; if two plates $\theta=\pm\alpha(t)$ are hinged at $r=0$ with viscous fluid in the gap $-\alpha(t)<\theta<\alpha(t)$, and if $\alpha(t)$ is slowly increased from a very small value, then the pressure $p(r,t)$ behaves like $\mu\,\dot{\alpha}(t) \log r$ which becomes negative as $r$ decreases towards zero, so that cavitation is inevitable in this region. This is of course a two-dimensional idealisation, but the same physical mechanism is presumably responsible for the cavitation bubbles appearing in the present experiment.  

We should note further that, as a consequence of the no-slip condition, a thin residual film is left on both plates in the regions invaded by the fingers. The thickness $s$ of these residual films  is presumably given by  $ s/h \sim \mu U /\gamma$, the result obtained  by \cite{B61} for a bubble rising in a capillary tube containing viscous liquid. In our present situation, with $\mu\sim 70\, \tn{kg} \tn{ s}^{-1}\tn{m}^{-1} $, $\gamma\sim 0.07$ kg s$^{-2}$ and the observed creeping velocity $U$ in the range $10^{-4}$--$10^{-5} $m\,s$^{-1}$, this gives $s/h\sim10^{-1}$--$10^{-2}$. When $h\sim 0.1$mm, the residual film thickness $s$ is therefore in the range 1--10 $\mu$m.

\subsection{Evolution of an annular drop}\label{Sec_annular}
Figure \ref{Fig_Annulus} shows an  evolution sequence when the initial shape of the drop [panel (a)] is as near to a uniform circular annulus as could be achieved by hand.   The curvature of the inner boundary of the annulus was inevitably not quite uniform, and where it was maximal a perturbation immediately developed, as anticipated by the stability analysis of \S\ref{Sec_destabilising_bubble}. This perturbation evolves into a secondary bubble [panel (b)] which is ejected, leaving the primary bubble [panel (c)] which deforms slowly to a distinctly non-circular shape  [panel(d)];  the outer boundary was only weakly perturbed from a circle,   The secondary bubble detaches from the primary bubble and erupts through the outer boundary [panels (d,e)] while a new (tertiary) bubble grows very slowly from the residual cusp on the primary bubble.  After about 45 minutes, the situation [panel(f)] appears to be quasi-static, and very small wrinkles appear on the outer boundary, probably associated with slight roughness of the two plates.  After a further hour, leverage was introduced at the lower right-hand corner, and fingering rapidly developed [panels (g-i)] over about one minute. A growing finger punctures the tertiary bubble [panel (g)] at the point marked by the arrow, providing a connected path from the primary bubble to the exterior. Fingers also erupt from the bubble [panel (j)] and encounter the `invading' fingers, forming a defensive ridge.  The pattern continues to evolve round the outer boundary [panel (k)], and cavitation bubbles may be detected in the `north-western' sector ahead of the advancing fronts. Panel [l] shows the ultimate fingering pattern just before rupture of the film when the plates separate completely; the ridge, which remains a prominent feature of this pattern, is breached only at the location of the puncture point.

\subsection{Merging of four drops}
We then carried out an experiment starting with four drops placed as near to the corners of a square as could be achieved, as shown in figure \ref{Fig_four_drops} [panel (a)]. As the upper plate descends, these drops expand and ultimately merge, trapping an air bubble  [panel (e)].  The initial drop contact [panel (c)] involves an instantaneous geometrical singularity at a time $\tau=\tau_{c}$, which is subsequently (i.e.~for  $\tau>\tau_{c}$) resolved by surface tension. [We should distinguish such a singularity from the more familiar finite-time singularities, as treated for example by \cite{E2015}, for which a singularity develops as $\tau\rightarrow\tau_{c}$ and is resolved by surface tension or some other mechanism just before the singularity occurs, i.e.~for $\tau<\tau_{c}$.]  \, After the bubble is trapped, it shows no tendency to become circular; on the contrary, where the curvature on the air bubble is maximal, a secondary bubble begins to pinch off [panel (g)], again leaving the very thin residual films of treacle on both plates. 

The formation of the secondary bubble is consistent with the qualitative description of the bubble instability in  \S\ref{Sec_destabilising_bubble} above:  the pressure contours are compressed where the bubble curvature is large, leading to amplification of the perturbation in this region.  The secondary bubble slowly propagates towards the outer boundary, separates from the primary bubble (a singular process analogous to the breaking of a viscous thread),  and ultimately erupts through this boundary [panel (i)] -- again a singular process.  Up to this point, the total volume of trapped air is constant,  but this falls very rapidly at the moment of eruption. At the same time, a tertiary bubble begins to form under the continuing influence of the pressure gradient; this suggests an iterative process, although much slower at each stage. 

\subsection{Contraction by levering}\label{Sec_levering}
Figure \ref{Fig_fingering} shows a continuation of the sequence of figure \ref{Fig_four_drops}, showing first [panels (a-c)] the eruption of the secondary bubble and the very slow separation of the tertiary bubble from the primary bubble which is now much reduced in size. Note that the outer boundary remains relatively smooth but does not become circular; this is because the internal instability has a persistent weak disturbing effect on the outer boundary [see the comment following (\ref{amplitude_ratios})]. In panel (d), levering has just commenced at the  lower right-hand corner of the plates, and  the first signs of the fingering instability are visible.  In panel (e), fingers erupting from the outer boundary of the drop are evident, and tip-splitting and side-branching of the invading fingers are well developed;  in panel (f)  a ridge develops between the invading and defending cohorts of fingers. In panel (g) the defending cohort, now in retreat, is punctured by an invading finger; the pressure is now atmospheric ($p_{a}$) throughout the fingers and the bubbles, but less than atmospheric in the remaining bulk of treacle; in panel (h) the invading  front of fingers advances round the boundary and cavitation bubbles appear ahead of the front.  Finally, panel (i) shows that fingering has again extended throughout the drop, to the extent that the treacle is almost entirely contained in the narrow `tree branches and twigs' that separate the fingers.  The ultimate pattern retains an imprint of the original trapped bubbles, in the pronounced pentagonal `ridge', still breached at only one point in the south-east sector.


\section{Conclusions}
The main  conclusions of this  investigation may be summarised as follows.\\
\noindent (i) A single drop of viscous fluid subject to squeezing under a constant force $F\!>\!0$ in a Hele-Shaw cell tends to adopt a circular form, with  radius $a(t)$ increasing like $t^{1/8}$ (the `one-eighth power law'); this is confirmed experimentally. \\
\noindent (ii)  An initially elliptic drop subject to similar squeezing tends to circular form; more generally, any perturbation of an expanding drop of  circular form  decreases algebraically in time,  i.e.~the circular form is stable.\\
\noindent  (iii) If the same drop is subject to contraction $(F\!<0)$, then it is unstable to a fingering disturbance whose scale is determined by surface tension; this fingering  establishes a lattice that ultimately extends throughout the fluid domain, before  rupture of the film  and complete separation of the plates. \\
\noindent (iv) Sufficiently large surface tension can lead to growth of $a(t)$ even when $F\!<0$, an important aspect of adhesion dynamics.\\
\noindent (v)  An annular drop subject to squeezing can also  follow a one-eighth power law expansion of both inner and outer boundaries; but this expansion is unstable, the instability growing predominantly from the inner boundary which shows no tendency to remain circular. \\
\noindent (vi) The instability manifests itself through ejection of a secondary bubble from the primary bubble leaving a residual layer on both plates whose thickness is determined by surface tension; this secondary bubble erupts through the outer boundary of the annular drop, and a tertiary bubble emerges from the residual cusp on the primary bubble; this process is 
extremely slow. \\
\noindent (vii)  When levering is applied to separate the plates, the fingering instability occurs at the outer boundary of the annulus, and rapidly spreads towards the inner boundary where much weaker fingers emerge; the bubble is punctured at a single  point by an invading finger, leading to equalisation of the pressure in the bubble with the external atmosphere.  It is like the puncturing of an inflated balloon. \\
\noindent (viii) As fingering continues to spread round the outer boundary, cavitation bubbles appear in the sector opposite the point of leverage, where the pressure would otherwise fall below the vapour pressure of the treacle; fingering ultimately extends over the whole fluid domain just before the final sudden rupture of the film. A ridge is formed where the invading cohort of fingers from the outer boundary impacts the defending cohort of fingers that emerge from the bubble. \\
\noindent (ix) A similar sequence of events occur when the air bubble is trapped by the expansion of four initially disjoint drops placed at the corners of a square; in this case the cusp singularities that occur when the drops make contact are resolved by surface tension as the drops continue to expand. Under subsequent leverage, the processes of fingering, puncturing, ridge formation and cavitation occur in similar manner in this case also.

\end{document}